\documentclass[a4paper,11pt]{article} 
\usepackage{amsmath}
\usepackage{graphicx}
\usepackage{amsfonts}
\usepackage{enumerate}
\usepackage{xcolor}
\usepackage{amsthm}
\usepackage{amssymb}
\usepackage{subfig}
\usepackage{caption}
\usepackage{listings}
\usepackage{bibunits}
\usepackage[greek,english]{babel}
\usepackage{wrapfig}
\usepackage{float}
\usepackage{url}

\theoremstyle{definition}

\theoremstyle{remark}

\usepackage{graphicx}
\usepackage[utf8]{inputenc}
\usepackage{gensymb}
\usepackage{natbib}
\usepackage{authblk}
\usepackage[left=3cm,right=3cm,top=3cm,bottom=3cm]{geometry}

\newcommand{\norm}{\mathrm{N}}
\newcommand{\gam}{\mathrm{Gam}}
\newcommand{\E}{\mathnormal{E}}
\newcommand{\Var}{\text{Var}}
\newcommand{\Cov}{\text{Cov}}

\newcommand{\diag}{\mathrm{diag}}

\def\T{{ \mathrm{\scriptscriptstyle T} }}
\renewcommand{\vec}[1]{\boldsymbol{#1}}
\newcommand{\vecn}[1]{\boldsymbol{#1}}
\newcommand{\matr}[1]{\mathnormal{#1}}
\newcommand{\matrn}[1]{\mathrm{#1}}
\newcommand{\VAR}[2]{\textsc{VAR}$_{#1}\text{(}{#2}\text{)}$} 
\newcommand{\AR}[1]{\textsc{AR}$\text{(}{#1}\text{)}$}

\title{Bayesian inference on the order of stationary vector autoregressions}
\date{}

\author[1]{Rachel~L.~Binks}
\author[2]{Sarah~E.~Heaps\thanks{Corresponding author: \textit{sarah.e.heaps@durham.ac.uk}}}
\author[3]{Mariella~Panagiotopoulou}
\author[3]{Yujiang~Wang}
\author[2]{Darren~J.~Wilkinson}
\affil[1]{School of Mathematics, Statistics and Physics, Newcastle University}
\affil[2]{Department of Mathematical Sciences, Durham University}
\affil[3]{School of Computing, Newcastle University}

\begin{document}

\maketitle

\begin{abstract}
Vector autoregressions (VARs) are a widely used tool for modelling multivariate time-series. It is common to assume a VAR is stationary; this can be enforced by imposing the stationarity condition which restricts the parameter space of the autoregressive coefficients to the stationary region. However, implementing this constraint is difficult due to the complex geometry of the stationary region. Fortunately, recent work has provided a solution for autoregressions of fixed order $p$ based on a reparameterization in terms of a set of interpretable and unconstrained transformed partial autocorrelation matrices. In this work, focus is placed on the difficult problem of allowing $p$ to be unknown, developing a prior and computational inference that takes full account of order uncertainty. Specifically, the multiplicative gamma process is used to build a prior which encourages increasing shrinkage of the partial autocorrelations with increasing lag. Identifying the lag beyond which the partial autocorrelations become equal to zero then determines $p$. Based on classic time-series theory, a principled choice of truncation criterion identifies whether a partial autocorrelation matrix is effectively zero. Posterior inference utilizes Hamiltonian Monte Carlo via Stan. The work is illustrated in a substantive application to neural activity data to investigate ultradian brain rhythms.
\end{abstract}

\section{\label{sec:intro}Introduction}

Vector autoregressive (VAR) processes are widely used to model multivariate time-series data in a variety of fields including neuroscience \citep[][]{chiang_bayesian_2016}, bioinformatics \citep[][]{jiang_microbial_2013,hannaford_sparse_2023}, macroeconomics \citep[][]{ECO-013}, and energy economics \citep[][]{heaps_gas_2020}. In an autoregression of order $p$, the random variable at time $t$ is conditionally independent of its values at lags $p+1, p+2, \ldots$ given observations at the preceding $p$ time points. Indeed the random variable at time $t$ can be expressed as a noisy linear combination of these $p$ values. The order of the autoregression is therefore intrinsic to the characterization of the joint process and plays a vital role in forecasting. However, its value is typically not known \textit{a priori}.

A common assumption when working with Gaussian time-series is that of stationarity, which posits that the means, variances and covariances of the process do not change over time. Since the overall level of many time-series exhibits periodic or systematic variation due to seasonality or time-trends, stationarity is often implausible as an assumption when modelling the raw data. However, stationary vector autoregressions frequently form the core building block of more sophisticated models, for example for differenced data in integrated models, for innovations from a time-varying mean in a time-series regression model or simply as components in state space models which are thought to be mean-reverting. From a practical perspective, enforcing stationarity prevents the predictive variance of the process from growing without bound into the future. This is often keenly motivated, for instance in applications where the goal is long-term forecasting or when modelling the dynamics of a linear system which is assumed to be in its equilibrium distribution. Moreover, stationarity admits various interpretations of the relationships between variables through the infinite-order moving average representation of the process, for example in Granger causality graphs or impulse response analysis.

Stationarity can be enforced by restricting the autoregressive coefficient matrices to lie within a constrained parameter space called the stationary region. However, the highly complex geometry of this region hampers the process of specifying a prior and subsequent computational inference. Fortunately, in recent work, \citet{heaps_enforcing_2022} solved the problem for vector autoregressions of fixed order by introducing an unconstrained and interpretable reparameterization of the stationary model. This is constructed by mapping the original model parameters to a set of partial autocorrelation matrices, which can be regarded as a vector analogue of the partial autocorrelation function of a univariate autoregression. A second transformation then scales the singular values of each of these partial autocorrelation matrices from $[0,1)$ to the positive real line. The transformed partial autocorrelation matrices are interpretable and allow specification of a prior which is invariant with respect to the order of the components in the observation vector. Markov chain Monte Carlo (MCMC) methods for computational inference need only operate over a Euclidean space, making implementation routine. However, a clear limitation of this work is that inference is conditional on a fixed order of the process, with no account for the uncertainty in its value.

Extending this work, we focus on development of a prior and associated procedures for computational inference which allow the order of stationary vector autoregressions to be learned. Although the general problem is hitherto unaddressed in the literature, it has been widely studied in the special case of univariate stationary autoregressions. In an approach which is similar in spirit to ours, \citet{barnett_bayesian_1996} enforce stationarity by reparameterizing the univariate model in terms of its partial autocorrelations, each with support restricted to $(-1, 1)$. A univariate stationary autoregression of order $p$ has a non-zero partial autocorrelation at lag $p$ and zero partial autocorrelations at all higher lags. Spike-and-slab priors, with an atom of probability at zero, are then chosen for each partial autocorrelation in a model with a large (maximum) value for $p$, so that identifying the lag of the final non-zero autocorrelation identifies the order of the process. \cite{vermaak_reversible_2004} use the same reparameterization of the model to enforce stationarity but frame the problem of order determination as a model selection problem and use reversible jump Markov chain Monte Carlo to learn the order of the process. In \cite{huerta_priors_1999}, stationarity is enforced through a different reparameterization of the univariate autoregression in terms of the reciprocal roots of its characteristic equation. Under this parameterization, the process is stationary if and only if the reciprocal roots have moduli less than 1 and the order of the process is determined by the number of reciprocal roots with non-zero modulus. Priors are assigned to the real and complex reciprocal roots with atoms of probability at moduli 0 in each case, thereby allowing inference on the model order. In the Bayesian framework, the only generalization of these ideas to stationary \emph{vector} autoregressions appears in \cite{huerta_structured_2006} who extend \cite{huerta_priors_1999} by considering a multivariate generalization of the characteristic equation. However, because this generalization is only available when the autoregressive coefficient matrices are diagonal, the approach is limited to the class of diagonal vector autoregressive processes. Other recent work which addresses the problem of order determination in vector autoregressions includes \cite{zhang_bayesian_2021} and \cite{fan_bayesian_2022} but their focus is on classes of rank-reduced models and stationarity is not enforced.

In this paper, we provide the first methodology for quantifying uncertainty in model order for the full class of stationary vector autoregressions. To this end, we enforce stationarity using the transformed partial autocorrelation parameterization of \citet{heaps_enforcing_2022}, exploiting a number of its properties to build a shrinkage prior for an overparameterized model. In particular, under this parameterization, the model of order $k$ is nested within the model of order $k+1$. We can therefore fit an overparameterized model with purposefully more lags than are required and construct a prior which increasingly shrinks the transformed partial autocorrelation matrices at higher lags towards zero. By identifying the lag beyond which the partial autocorrelations become essentially equal to zero, we learn about the order of the process. The interpretability of the reparameterized model allows classical theory on the sampling distribution of the partial autocorrelation function to inform specification of the shrinkage prior and subsequent decision-making about whether a partial autocorrelation matrix is effectively zero. This is based on a truncation criterion for the max norm of the matrix which is completely general and could be applied with any prior for an overparameterized stationary model. We also provide code for implementation of computational inference via Stan thereby facilitating straightforward use by a variety of practitioners across the spectrum of fields which rely on vector autoregressions for modelling and forecasting applications.

The remainder of the paper is structured as follows. In Section~\ref{sec:VARs} we discuss the core reparameterization of stationary vector autoregressive models in terms of transformed partial autocorrelation matrices. In Section~\ref{sec:prior} we discuss the prior distribution assigned to the unknowns in our reparameterized model. Section~\ref{sec:posterior_inference} considers posterior inference and the use of a truncation criterion to determine model order. In Section~\ref{sec:simulation_studies} we apply our model and inferential procedures in a set of simulation experiments before considering a substantive application to neural activity data in Section~\ref{sec:application}. Finally Section~\ref{sec:discussion} provides some concluding remarks.

\section{\label{sec:VARs}Stationary vector autoregressions}

\subsection{Vector autoregressions}

Without loss of generality, suppose that the $m$-variate process $\{ \vec{y}_t \}$ can be modelled as a zero-mean vector autoregression of order $p$, denoted \VAR{m}{p},
\begin{equation}\label{eq:var}
\vec{y}_t = \matr{\phi}_1 \vec{y}_{t-1} + \ldots + \matr{\phi}_p \vec{y}_{t-p} + \vec{\epsilon}_t,
\end{equation}
where the errors $\vec{\epsilon}_t$ form a sequence of uncorrelated, zero-mean multivariate normal random vectors, $\vec{\epsilon}_t \sim \norm_m(\vecn{0}, \matr{\Sigma})$. The continuous model parameters therefore comprise the autoregressive coefficient matrices $\matr{\phi}_i \in M_{m\times m}(\mathbb{R})$, $i=1,\ldots,p$, which are denoted collectively as $\matr{\Phi} \in M_{m\times m}(\mathbb{R})^p$, and the error variance matrix $\matr{\Sigma} \in \mathcal{S}_m^+$, where $M_{m\times n}(V)$ and $\mathcal{S}_m^+$ denote the space of $m \times n$ matrices with entries in $V$ and the space of $m \times m$ symmetric, positive definite matrices, respectively. Defining $B$ as the backshift operator, such that $B^k \vec{y}_t = \vec{y}_{t-k}$, it is common to express~\eqref{eq:var} as
\begin{equation*}
\vec{\epsilon}_t = (\matr{I}_m - \matr{\phi}_1 B - \ldots - \matr{\phi}_p B^p) \vec{y}_t = \phi(B) \vec{y}_t,
\end{equation*}
in which $\matr{I}_m$ is the $m \times m$ identity matrix and $\phi(u) = (\matr{I}_m - \matr{\phi}_1 u - \ldots - \matr{\phi}_p u^p)$, $u \in \mathbb{C}$, is referred to as the characteristic polynomial. A vector autoregression is stable if and only if all the roots of $\det\{ \phi(u) \} = 0$ lie outside the unit circle. Since all stable processes are stationary, and unstable stationary processes are not generally of interest, this is often referred to as the stationarity condition for $\matr{\Phi}$ and the subset of $M_{m\times m}(\mathbb{R})^p$ over which the condition is satisfied is referred to as the stationary region, denoted $\mathcal{C}_{p,m}$.

\subsection{\label{subsec:reparam}Reparameterization over the stationary region}
As illustrated in Section S1 of the Supplementary Materials \citep[][]{supplMat}, the stationary region $\mathcal{C}_{p,m}$ has a complex geometry, especially when $p$ or $m$ is large. With no standard distributions over $\mathcal{C}_{p,m}$, this complicates the process of specifying a prior that conveys meaningful information, for example, concerning the relative sizes of the autocorrelations at different lags. Moreover, it is difficult to design an efficient MCMC sampler which targets a distribution with support constrained to $\mathcal{C}_{p,m}$. In recent work, \citet{heaps_enforcing_2022} proposes a solution which addresses both issues, reparameterizing the model over the stationary region in terms of a set of interpretable, unconstrained parameters. The reparameterization involves two bijective mappings. First, the original model parameters $(\matr{\Sigma}, \matr{\Phi}) \in \mathcal{S}_m^+ \times \mathcal{C}_{p,m}$ are mapped to a new parameter set $\{\matr{\Sigma}, (\matr{P}_1, \ldots, \matr{P}_p) \} \in \mathcal{S}_m^+ \times \mathcal{V}^p$ in which $\mathcal{V}$ denotes the subset of matrices in $M_{m\times m}(\mathbb{R})$ whose singular values are less than one. The matrix $\matr{P}_{s+1}$ is referred to as the $(s+1)$-th partial autocorrelation matrix. It is defined as the conditional cross-covariance matrix between $\vec{y}_{t+1}$ and $\vec{y}_{t-s}$ given $\vec{y}_t, \ldots, \vec{y}_{t-s+1}$ which has been standardized through 
\begin{equation*}
\matr{P}_{s+1} = \matr{\Sigma}_s^{-1/2} \Cov( \vec{y}_{t+1}, \vec{y}_{t-s} | \vec{y}_t, \ldots, \vec{y}_{t-s+1}) \matr{\Sigma}^{* \, -1/2}_s,
\end{equation*}
$s=0,\ldots,p-1$, in which $\matr{\Sigma}_s$ and $\matr{\Sigma}_s^*$ are the conditional variances
\begin{equation*}
\matr{\Sigma}_s = \Var(\vec{y}_{t+1} | \vec{y}_t, \ldots, \vec{y}_{t-s+1}) \quad \text{and} \quad \matr{\Sigma}_s^* = \Var(\vec{y}_{t-s} | \vec{y}_{t-s+1}, \ldots, \vec{y}_t)
\end{equation*}
and $\matr{\Sigma}^{1/2}$ denotes the symmetric matrix-square-root. Full details of the mapping and its inverse, which proceed by recursion, are described in \citet{heaps_enforcing_2022}. The steps are also outlined in Section S2 of the Supplementary Materials \citep[][]{supplMat}, with a numerical example for the simple case where $m=2$ and $p=2$. 

A second transformation then maps each partial autocorrelation matrix $\matr{P} \in \mathcal{V}$ to an unconstrained square matrix $\matr{A} \in M_{m \times m}(\mathbb{R})$ through
\begin{equation*}
\matr{A} = (\matr{I}_m - \matr{P}\matr{P}^\T)^{-1/2} \matr{P}.
\end{equation*}
Denoting the singular value decomposition of $\matr{P}$ by $\matr{P}=\matr{U}\diag(r_1, \ldots, r_m)\matr{V}^\T$ in which the singular values satisfy $1 > r_1 \ge r_2 \ge \cdots \ge r_m \ge 0$, the corresponding factorization of $\matr{A}$ is given by $\matr{A}=\matr{U}\diag(\tilde{r}_1, \ldots, \tilde{r}_m)\matr{V}^\T$ where $\tilde{r}_i = r_i / (1 - r_i^2)^{1/2} \ge 0$, $i=1,\ldots,m$. Therefore the second transformation can be regarded as an orientation-preserving mapping which simply scales the singular values of $\matr{P}$ from $[0,1)$ to the positive real line.

\section{\label{sec:prior}Prior distribution}

\subsection{Shrinkage prior for transformed partial autocorrelations}

The relationship between the singular value decompositions of $\matr{P}$ and $\matr{A}$, described in the previous section, has two important implications which we exploit in building our prior. First, the spectral norms of $\matr{P}$ and $\matr{A}$, $r_1 = \| \matr{P} \|_2$ and $\tilde{r}_1 = \| \matr{A} \|_2$, are clearly related through the monotonic mapping: $\tilde{r}_1 = r_1 / (1 - r_1^2)^{1/2}$. The relative sizes of the unconstrained parameters $\matr{A}_s$ across lags $s=1,\ldots,p$ therefore relate directly to the relative sizes of the partial autocorrelation matrices $\matr{P}_s$ across lags. Second, $\matr{P}=\matrn{0}_m$ if and only if $\matr{A}=\matrn{0}_m$ in which $\matrn{0}_m$ denotes the $m \times m$ matrix of zeros. It follows from the definition of the partial autocorrelation matrices that for $k < p$, $\matr{P}_k \ne \matrn{0}_m$ and $\matr{P}_{k+s} = \matrn{0}_m$ for $s=1,\ldots,p-k$ if and only if $\matr{\phi}_k \ne \matrn{0}_m$ and $\matr{\phi}_{k+s} = \matrn{0}_m$ for $s=1,\ldots,p-k$. The order of a \VAR{m}{p} process is therefore $k < p$ if and only if $\matr{A}_k \ne \matrn{0}_m$ and $\matr{A}_{k+s} = \matrn{0}_m$ for $s=1,\ldots,p-k$. Under the unconstrained parameterization, it follows that the model of order $k < p$ is nested within the model of order $k + 1$. 

A common Bayesian approach to quantifying uncertainty on the dimension of nested models is to fit an overparameterized model with purposefully more components than are required. By using a shrinkage prior, components that are shrunk enough to be deemed negligible in the likelihood can then be discarded. Consequently, inference on both the continuous model parameters and the model dimension are available from a single within-model MCMC sampler, without recourse to transdimensional MCMC; see, for instance, \citet{rousseau_overfitting_2011} and \citet{bhattacharya_sparse_2011} in the context of mixture and factor models, respectively. We can therefore borrow ideas from this literature by adopting a shrinkage prior for $\matr{A}_1, \ldots, \matr{A}_{p_{\text{max}}}$ with a large value for $p_{\text{max}}$. By identifying the lag beyond which the $\matr{A}_s$ can be taken as zero matrices, we therefore learn about the order $p$ of the process. Moreover, we can convey the very reasonable idea that the partial autocorrelations at higher lags are likely to be smaller than those at lower lags by choosing a shrinkage prior for the $\matr{A}_s$, $s=1,\ldots,p_{\text{max}}$, whose degree of shrinkage increases with the lag $s$.

A popular increasing shrinkage prior is the multiplicative gamma process (MGP) \citep[][]{bhattacharya_sparse_2011} originally developed as a structured sequence of global-local shrinkage priors for the loadings matrix in infinite factor models. Denoting the $(i,j)$th element in $\matr{A}_s$ by $a_{s,ij}$ we adopt a prior of this form by choosing
\begin{equation*}
a_{s,ij} | \lambda_{s,ij},\tau_s \sim \norm(0,\lambda_{s,ij}^{-1}\tau_s^{-1}),
\end{equation*}
independently for $i,j=1,\ldots,m$, $s=1,\ldots,p_{\text{max}}$, where the local precision parameters at lag $s$ are assigned the prior
\begin{equation*}
\lambda_{s,ij} \sim \gam(a/2, a/2),
\end{equation*}
independently for $i,j=1,\ldots,m$, $s=1,\ldots,p_{\text{max}}$, and the global precision parameter at lag $s$ is constructed as
\begin{equation*}
\tau_s = \prod_{k=1}^s \delta_{k}, \quad \delta_{1} \sim \gam(a_1, 1), \quad \delta_{k} \sim \gam(a_2, 1), \quad k \ge 2,
\end{equation*}
in which the $\delta_{k}$ are independent. The global precisions $\tau_s$ are therefore a cumulative product of gamma random variables whose prior expectation $\E(\tau_s)$ increases with $s$ when $a_2 > 1$. Guidelines on the choice of hyperparameter $a_1$ and $a_2$ can be found in \citet{durante_note_2017} who presents a numerical method for checking that the global variances $\theta_s = 1 / \tau_s$ are stochastically decreasing in $s$ near zero, that is, $\Pr\{\theta_s \in (0,\theta]\}$ is non-decreasing in $s$ for any $\theta$ in a small neighbourhood of zero. 

The multiplicative gamma process prior does not place any mass at zero and so none of the $\matr{A}_s$, and hence $\matr{P}_s$, matrices are shrunk exactly to zero. We define the \emph{effective order} $p^{\ast}$ of the model as the value of $s \le p_{\text{max}}$ such that $\matr{P}_{s}$ fails a criterion for truncation to zero when $s=p^{\ast}$ but passes for $s=p^{\ast}+1,\ldots,p_{\text{max}}$. Applying the truncation criterion to the standardized $\matr{P}_s$ matrices, rather than the unconstrained $\matr{A}_s$ matrices, allows classical theory from univariate time-series analysis to inform our judgement in a manner which is robust with respect to the scale of the data as well as its length and dimension $m$. Further details on the truncation criterion are provided in Section~\ref{subsec:truncation_criterion}. It is worth emphasising that although our prior encourages increasing shrinkage of the $\matr{P}_s$ matrices towards zero as the lag increases, this ordering does not have to be replicated in the posterior if it is not supported by the data. Further, although by definition $\matr{P}_{p^{\ast}}$ is not truncated to zero for $p^{\ast}>1$, any matrix $\matr{P}_s$, where $s \in \{1,\ldots, p^{\ast}-1\}$, can be truncated to zero.

An alternative inferential approach would have been to adopt a spike-and-slab prior for each unconstrained $\matr{A}_s$ matrix. This is the basis of the Bayesian stochastic search procedure that is often used for individual coefficients in the analysis of autoregressions \citep[][]{GSN08,SCL06}. An increasing shrinkage prior within this class is the cumulative shrinkage process (CUSP) \citep[][]{legramanti_bayesian_2020}. However, our experience working with this prior suggests that, though sensible posterior inferences can be obtained in the analysis of simulated data, inference on the model order is very sensitive to the choice of prior hyperparameters in analyses involving real data. This suggests a lack of robustness to the kind of model misspecification that is inevitable in analyses of real time-series. Further discussion of these approaches can be found in Chapter 5 of \citet{Bin24}.

\subsection{Joint prior}

Denoting the collection of unknown hyperparameters in the multiplicative gamma process prior by $\vec{\vartheta}$, we adopt an overall prior specification of the form
\begin{equation}\label{eq:prior}
\pi(\matr{\Sigma}, \matr{A}_1, \ldots, \matr{A}_{p_{\text{max}}}, \vec{\vartheta}) = \pi(\matr{\Sigma}) \pi(\vec{\vartheta}) \prod_{s=1}^{p_{\text{max}}} \pi(\matr{A}_s | \vec{\vartheta}).
\end{equation}
Various options are available for the error variance matrix $\matr{\Sigma}$ and distributions which offer the property of invariance with respect to the order of the variables in the observation vector are discussed in \citet{heaps_enforcing_2022}. In the applications in this paper, we use one such distribution, taking $\matr{\Sigma}$ to be inverse Wishart, with a scale matrix that has a common element on the diagonal and a common element off the diagonal.

\section{\label{sec:posterior_inference}Posterior inference}


\subsection{Posterior distribution}

For $i \le j$, denote by $\vec{y}_{i:j}$ the time-series $\vec{y}_i, \ldots, \vec{y}_j$. The likelihood for a series of $n$ observations, $\vec{y}_{1:n}$, from a zero-mean \VAR{m}{p_{\text{max}}} process can be expressed as
\begin{equation*}
p(\vec{y}_{1:n} \mid \matr{\Sigma}, \matr{\Phi}) = p(\vec{y}_{1:p_{\text{max}}} \mid \matr{\Sigma}, \matr{\Phi})  \prod_{t=p_{\text{max}}+1}^n p(\vec{y}_t \mid \vec{y}_{(t-p_{\text{max}}):(t-1)}, \matr{\Sigma}, \matr{\Phi})
\end{equation*}
in which $\vec{Y}_t \mid \vec{y}_{(t-p_{\text{max}}):(t-1)}, \matr{\Sigma}, \matr{\Phi} \sim \norm_{m}\left(\sum_{i=1}^{p_{\text{max}}} \matr{\phi}_i \vec{y}_{t-i} \, , \, \matr{\Sigma}\right)$ and the initial distribution is $(\vec{Y}_1^\T, \ldots, \vec{Y}^\T_{p_{\text{max}}})^\T \mid \matr{\Sigma}, \matr{\Phi} \sim \norm_{mp_{\text{max}}}(\vecn{0}, \matr{G})$. Here $\matr{G}$ is given by
\begin{equation*}
    \matr{G} = \begin{pmatrix}
    \matr{\Gamma}_0&\matr{\Gamma}_1&\cdots&\matr{\Gamma}_{p_{\text{max}}-1}\\
    \matr{\Gamma}_1^\T&\matr{\Gamma}_0&\cdots&\matr{\Gamma}_{p_{\text{max}}-2}\\
    \vdots&\vdots&\ddots&\vdots\\
    \matr{\Gamma}_{p_{\text{max}}-1}^\T&\matr{\Gamma}_{p_{\text{max}}-2}^\T&\cdots&\matr{\Gamma}_0\\
    \end{pmatrix},
\end{equation*}
where the matrices $\matr{\Gamma}_s=\Cov(\vec{y}_t, \vec{y}_{t+s})$, $s=0,\ldots,p_{\text{max}}-1$, are available as by-products of the recursive mapping between the partial autocorrelation matrices and the original model parameters.

Regarding the likelihood as a function of the new parameters and combining it with the prior~\eqref{eq:prior} via Bayes theorem yields the posterior distribution as
\begin{equation}\label{eq:posterior}
\pi(\matr{\Sigma}, \matr{A}_1, \ldots, \matr{A}_{p_{\text{max}}}, \vec{\vartheta} \mid \vec{y}_{1:n}) \propto \pi(\matr{\Sigma}) \pi(\vec{\vartheta}) \prod_{s=1}^{p_{\text{max}}} \pi( \matr{A}_s \mid \vec{\vartheta} ) p(\vec{y}_{1:n} \mid \matr{\Sigma}, \matr{A}_1, \ldots, \matr{A}_{p_{\text{max}}}).
\end{equation}
As explained in \citet{heaps_enforcing_2022}, the posterior distribution is a complicated function of the $\matr{A}_s$, making it ill-suited to computational inference based on Gibbs sampling. Rather than appealing to conditional independence structure in the posterior for one-at-a-time parameter updates, Hamiltonian Monte Carlo (HMC) \citep[][]{neal_mcmc_2011} uses information on the slope of the logarithm of the posterior density to generate global proposals that update all parameters simultaneously. We have found it efficient in sampling from the posterior~\eqref{eq:posterior} and use \texttt{cmdstanr} \citep[][]{cmdstanr_manual_2021}, a lightweight R interface to the Stan software \citep[][]{carpenter_stan_2017}, to implement the HMC algorithm. Stan requires users to write a program in the probabilistic Stan modelling language, the role of which is to provide instructions for computing the logarithm of the kernel of the posterior density function. The Stan software then automatically sets up a Markov chain simulation to sample from the resulting posterior. This includes calculation of the gradient of the logarithm of the posterior density, random initialization of the chains, and the tuning of the sampler.

\subsection{\label{subsec:truncation_criterion}Truncation criterion}

Following \citet{bhattacharya_sparse_2011}, we choose to truncate $\matr{P}_s$ to a zero matrix if the absolute values of all of its elements lie below some threshold, say $\varepsilon$. In classical time-series analysis, the partial autocorrelation plot, with its associated confidence intervals, plays an important role in the choice of order for a univariate autoregression. Under the hypothesis that the process is \AR{p}, the estimators for the partial autocorrelations of order $p+1,p+2,\ldots$ based on a sample of size $n$ are approximately independent with mean equal to zero and variance equal to $1/n$. As a guide, we therefore approximate the posterior for the $m^2$ components $p_{s,ij}$ of $\matr{P}_s$ under this hypothesis as independent $\norm(0, 1/n)$ random variables and then compute the quantile $q_{m,n}(\beta)$ such that $\Pr\{\max_{i,j} | p_{s,ij} | < q_{m,n}(\beta) \} = \beta$ for some large value of $\beta$ and set the threshold 
\begin{equation}\label{eq:truncation_criterion}
\varepsilon=q_{m,n}(\beta)=\Phi^{-1}\{ (\beta^{1/m^2}+1) / 2 \} / \sqrt{n}. 
\end{equation}
By choosing the threshold in this way, we account for both the length and dimension of the data, in addition to operating on a parameter which is unit-free. We typically set $\beta=0.99$, however, as discussed further in Section~\ref{sec:simulation_studies}, the posterior for the effective order $p^*$ is largely insensitive to sensible variation in the value of $\beta$, for example, from $\beta=0.95$ to $\beta=0.999$. 

For each draw from the posterior, we can apply this criterion to determine the effective order $p^*$ of the process. This can be summarized to yield a numerical approximation of the posterior for $p^*$ which provides a proxy for the posterior for $p$. We emphasize that this truncation criterion does not depend on the choice of prior and could be applied in conjunction with any prior for an overparameterized stationary vector autoregression in order to learn the order of the process.

\section{\label{sec:simulation_studies}Simulation experiments}

Consider the idealized setting in which we know that the data were generated from a stationary vector autoregression of known order, $p$. In order to explore the behaviour of the posterior distribution for $p^*$ in this context, we carried out simulation experiments that considered data generated from processes whose orders took various values. Our choice of truncation criterion makes allowance for the dimension of the observation vector $m$ and the length of the time-series $n$. We might therefore expect some degree of robustness in the more challenging inferential situations when $n$ is small or, in particular, when $m$ is large. This was investigated by considering simulations under a variety of values of $m$ and $n$.

\begin{figure}[!tb]
\begin{center}
\includegraphics[width=1\textwidth]{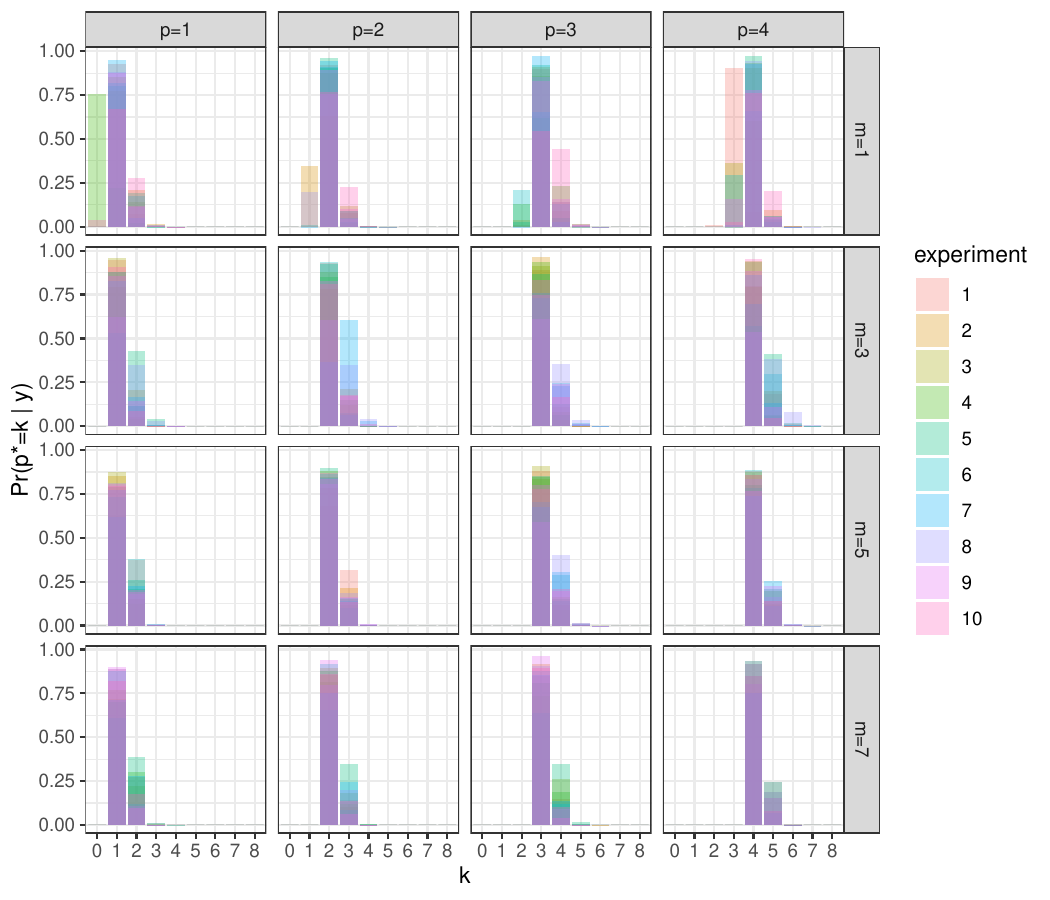}
\caption{Overlaid posterior mass functions for the effective order $p^*$ from 10 experiments for each combination of $m \in \{1,3,5,7\}$ and $p \in \{1,2,3,4\}$, with $n=1000$.}
\label{fig:post_mass_simulation}
\end{center}
\end{figure}

For each $m \in \{1, 3, 5, 7\}$ and $p \in \{1, 2, 3, 4 \}$ we simulated ten sets of $m \times m$ matrices $\matr{A}_1, \ldots, \matr{A}_p$ with elements sampled independently from a standard normal distribution. Taking the error variance matrix to be $\matr{\Sigma}=\matr{I}_m$, these were used to simulate ten stationary \VAR{m}{p} processes of length $n = 1000$. Conditional on each data set, we then generated samples from the posterior distribution using Stan, as described in Section \ref{sec:posterior_inference}, setting the maximum possible order as $p_{\max} = 8$. Values for the other hyperparameters in the prior are detailed in Section S3.1 of the Supplementary Materials \citep[][]{supplMat}. In all cases, we used four chains each with 1,000 iterations of warm-up followed by 4,000 sampling iterations. Using the truncation criterion with $\beta = 0.99$, we calculated the limits $q_{m,n}(\beta)$ as 0.081, 0.103, 0.112 and 0.117 for $m=1,$ 3, 5, and 7 respectively, and obtained a posterior mass function for the effective order $p^*$ of each process. The posterior mass functions are summarized in Figure \ref{fig:post_mass_simulation} across all simulation experiments. For a given $(m, p)$, the posteriors for the ten data sets are presented as a collection of overlaid bar charts. In all but three (of 160) cases, the true order $p$ of the process is the mode in the posterior for $p^*$, with considerable posterior support. The results are similar across different values of $m$ and $p$, suggesting robustness to the dimension of the data through our choice of truncation criterion.

\begin{figure}[!tb]
\begin{center}
\includegraphics[width=1\textwidth]{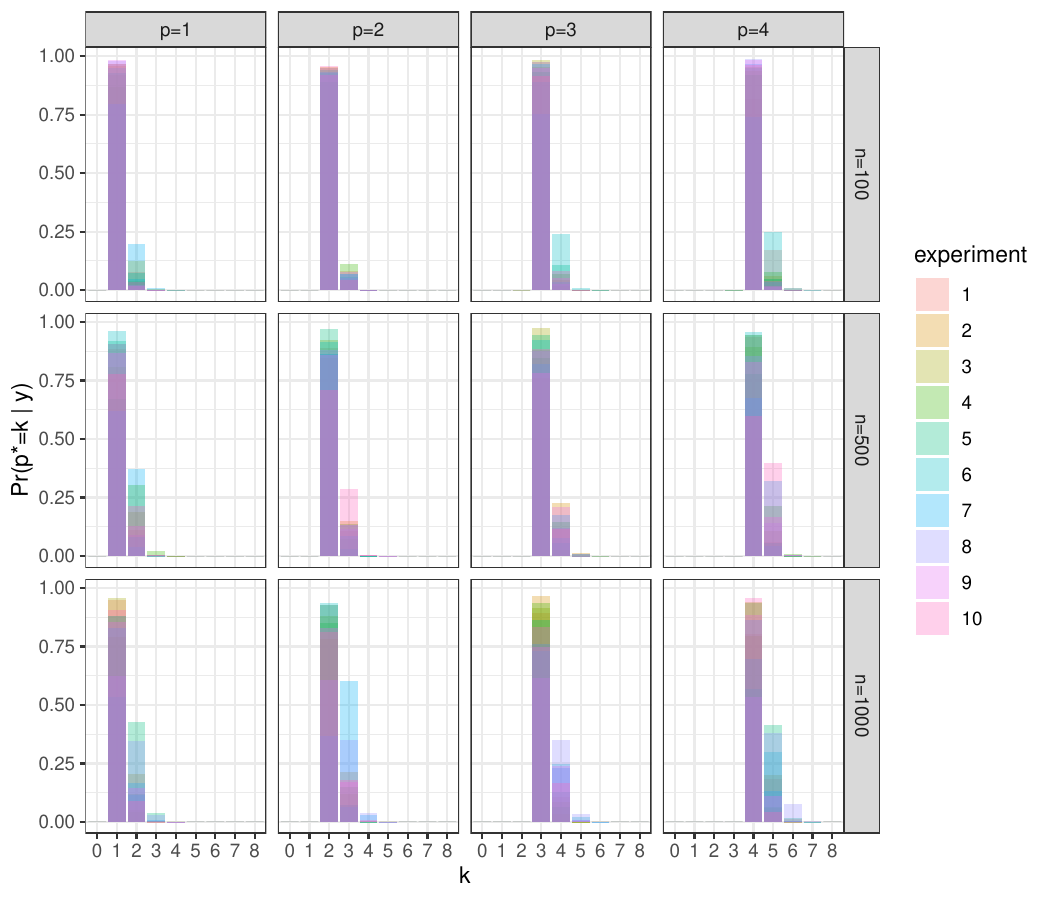}
\caption{Overlaid posterior mass functions for the effective order $p^*$ from 10 experiments for each combination of $n \in \{100, 500, 1000\}$ and $p \in \{1,2,3,4\}$, with $m=3$.}
\label{fig:post_mass_changing_n}
\end{center}
\end{figure}

Fixing $m=3$, considering $p \in \{1,2,3,4\}$ and using the same ten sets of matrices $\matr{A}_1, \ldots, \matr{A}_p$ as in the previous experiment, we then simulated ten stationary \VAR{m}{p} processes of length $n = 100$ and another ten of length $n = 500$, facilitating comparison across $n \in \{ 100, 500, 1000\}$. Retaining the same prior specification in the new experiments, we fit the model using HMC via Stan, as discussed above. Again, using the truncation criterion with $\beta = 0.99$ led to limits $q_{m,n}(\beta)$ equal to 0.326, 0.146 and 0.103 for $n = 100$, 500 and 1000, respectively. This yielded the posterior mass functions for $p^*$ which are displayed in Figure \ref{fig:post_mass_changing_n}. In all but one experiment for the different values of $n$, the posterior mode for the effective order $p^*$ recovers the true order $p$ of the process, again, with considerable support. The similarity in the behaviour of the posterior for all values of $n$  suggests robustness through the choice of truncation criteria, even for short time-series.

In order to assess sensitivity to our choice of truncation criterion we considered a selection of other choices for the probability $\beta$ in~\eqref{eq:truncation_criterion}. The results, presented in full in Section S3.2 of the Supplementary Materials \citep[][]{supplMat}, revealed that the posterior for the effective order $p^*$ of the process was largely insensitive to variation in the value of $\beta$ over a sensible range, with the posterior mode remaining equal to the known order of the process and an increasing degree of concentration about the mode as $\beta$ increased. Indeed, posterior mass only started to concentrate above the known order when the value of $\beta$ was purposely chosen to be unreasonably small, leading to overestimation of $p$. These observations held across all values of $m$, $n$ and $p$ which provides further empirical evidence of the utility of our principled approach for selecting the truncation criterion.

\section{\label{sec:application}Application: understanding brain rhythms}

\subsection{Background}
As an example application, we will apply our model and inferential procedures to a dataset of long-term intracranial EEG recordings to understand biological rhythms in the brain. Biological rhythms on ultradian, circadian, and longer timescales have been demonstrated in human physiology but particularly the ultradian rhythms remain elusive in mechanism and function in the brain \citep[][]{goh_episodic_2019-1, lloyd_occurrence_1991}. Multiple lines of evidence suggest that some prominent ultradian rhythms exist in brain activity as measured by EEG \citep[][]{hayashi_ultradian_1994, panagiotopoulou_fluctuations_2022}, and may be related to rest-activity cycles, or even modulate disease symptoms.

In this exploratory application we investigate the properties that such ultradian biological rhythms may display in human brain activity. We use band power in two common frequency bands (delta and beta) as our features of interest.

\subsection{Data preprocessing}

Intracranial EEG recordings are considered from four subjects with refractory focal epilepsy from the University College London Hospital (UCLH). We give the individuals the anonymous identities of A, B, C and D. The nature of the recording was chosen for its high signal-to-noise ratio without the need for extensive artefact detection and removal. 

Firstly, we divided each subject's iEEG data into non-overlapping, consecutive segments of length 30 seconds. All channels within each segment were re-referenced to a common average reference. In the common average calculation, channels with extreme amplitude values were excluded. A notch filter was then applied at $50$~Hz for each 30 second time window to remove line noise, after which the time windows were band-pass filtered from $0.5-80$~Hz using a fourth order zero-phase Butterworth filter (second order forward and backward filter applied) and downsampled to $200$~Hz. 

Next, the iEEG data were decomposed into commonly studied frequency bands, including delta and beta \citep[][]{taylor_normative_2022}. We calculated the iEEG band power for each 30 second segment for all channels in two frequency bands ($\delta:~1-4$~Hz, $\beta:~13-30$~Hz) using Welch's method with three-second non-overlapping windows. After taking logarithms to base 10 of the band power recordings in each channel, the channels were averaged into the brain regions from which they were recorded based on the Desikan-Killiany atlas; see \citet{wang_temporal_2023} for further details. The number of brain regions varied between individuals, with $m = 8$, 8, 9 and 13 for individuals A, B, C and D respectively. Finally, the data were mean-centered prior to analysis.

For each individual we analysed the longest possible contiguous time period of their band power time-series for which graphical interrogation of the data suggested stationarity was a plausible assumption; therefore the length of the recording chosen for further analysis varied across subjects. The number of observations in the recordings used were $n= 685$, 622, 651 and 231 for individuals A, B, C and D respectively, equivalent to 5.7, 5.175, 5.417 and 1.917 hours. These recordings were obtained during day-time hours.
We apply our model and inferential procedures to the time-series of both the beta and delta band power values for each individual, with a maximum order of $p_{\max} = 8$. The choices of hyperparameters in the prior are provided in Section S4.1 of the Supplementary Materials \citep[][]{supplMat}.

\subsection{MCMC implementation}
For each individual and frequency band, we used the \texttt{cmdstanr} interface to the Stan software (version 2.29.2) to run 4 chains, initialized at different starting points. The chains were run for 17,000 iterations, discarding the first 1,000 as burn-in, and thinning to retain every fourth draw in order to reduce computational overheads. The usual graphical and numerical diagnostics gave no evidence of any lack of convergence and, after pooling the chains, the effective sample size was at least 2,023 for every model parameter. Using a CentOS Linux 7 (Core) 64bit operating system with an Intel Xeon E5-2699 v4 processor (2.2 GHz), the average run time across chains for the beta frequency band in individual A was 31.3 hours, with similar run times obtained for the other data sets.

\subsection{Order determination}
\begin{figure}[!tb]
\begin{center}
\includegraphics[width=1\textwidth]{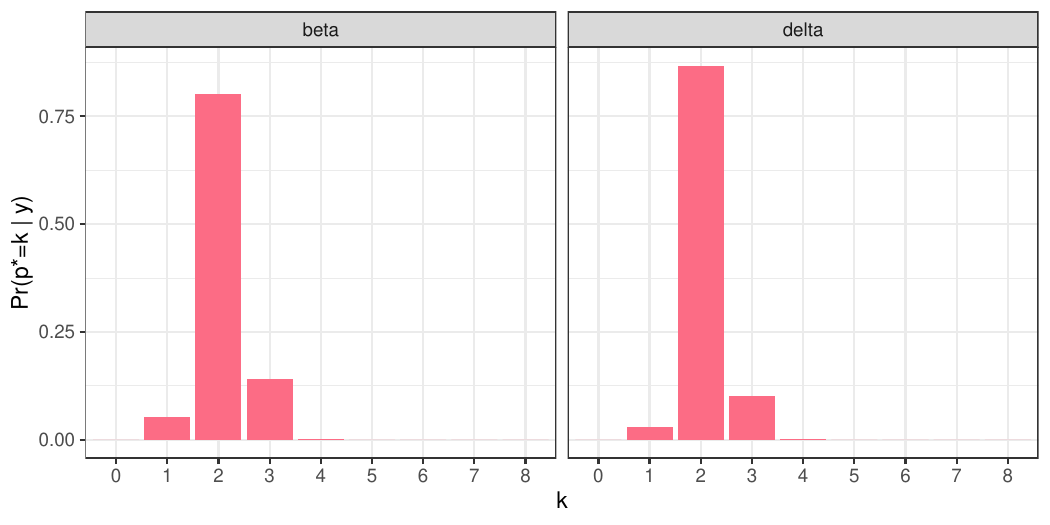}
\caption{Posterior mass function for the order of process for the beta (left) and delta (right) series for individual A.}
\label{fig:1110_day_pmf}
\end{center}
\end{figure}

For each of the individuals, the posterior distributions for $p^*$ for both the beta and delta series were calculated using the truncation criteria described in Section~\ref{subsec:truncation_criterion} with $\beta = 0.99$. For example, in Figure \ref{fig:1110_day_pmf}, the posterior mass functions for the data pertaining to individual A are shown. For both the delta and beta series, the posterior mode is 2, with posterior support exceeding $2/3$. Similar results were obtained using the commonly adopted frequentist strategy of selecting model order based on the Akaike's Information Criterion (AIC) or Schwartz Criterion (SC) \citep[][Chapter 4]{Lut06}, with the AIC selecting an order of 2 for the data in the beta and delta bands and the SC selecting 1 in each case. These results are quantitatively similar across all individuals, with figures provided in Section S4.2 of the Supplementary Materials \citep[][]{supplMat}, possibly indicating similar generative processes for their ultradian rhythms. 

\subsection{Granger causality}
\begin{figure}[!tb]
\begin{center}
\includegraphics[width=1\textwidth]{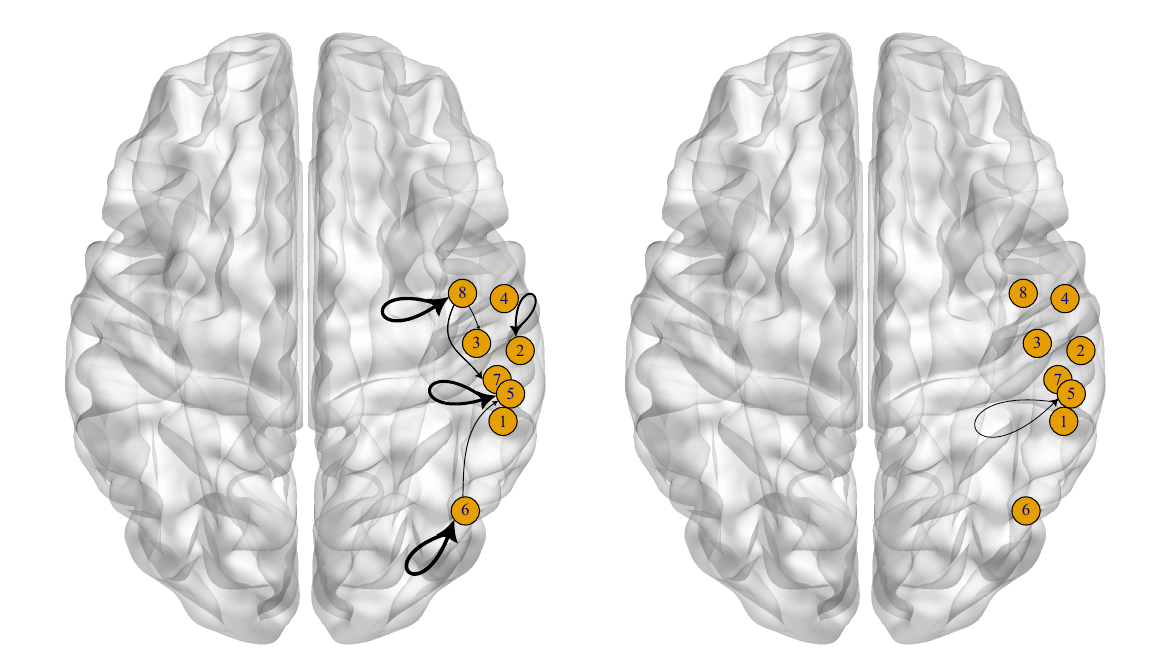}
\caption{Granger causality plots of the posterior mean of the autoregressive coefficient matrices for the VAR process of individual A in the beta band at lag 1 (left) and lag 2 (right), overlaid on glass brains showing the locations of the regions. Region names: 1 - r.bankssts, 2 - r.middletemporal, 3 - r.postcentral,  4 - r.superiortemporal, 5 - r.supramarginal, 6 - r.inferiorparietal, 7 - r.inferiortemporal, 8 - r.precentral.}

\label{fig:beta_1110_day_GC}
\end{center}
\end{figure}

\begin{figure}[!tb]
\begin{center}
\includegraphics[width=1\textwidth]{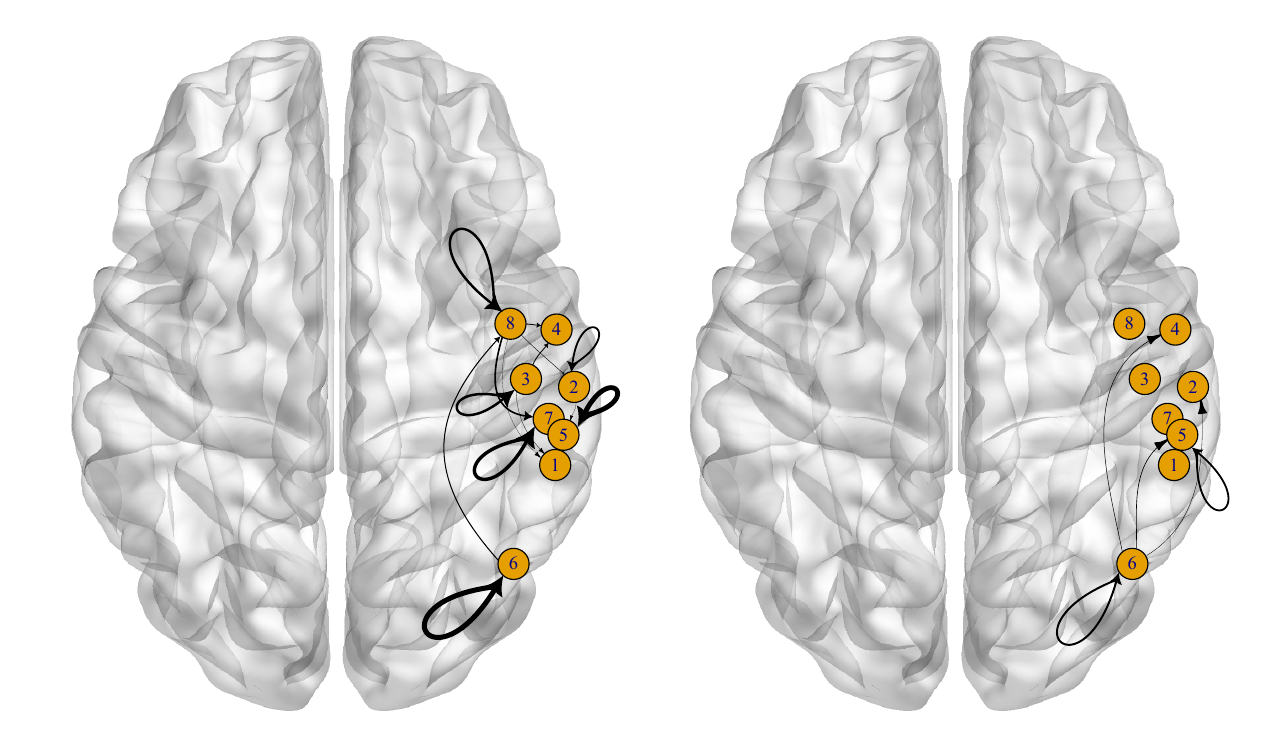}
\caption{Granger causality plots of the posterior mean of the autoregressive coefficient matrices for the VAR process of individual A in the delta band at lag 1 (left) and lag 2 (right), overlaid on glass brains showing the locations of the regions. Region names: 1 - r.bankssts, 2 - r.middletemporal, 3 - r.postcentral,  4 - r.superiortemporal, 5 - r.supramarginal, 6 - r.inferiorparietal, 7 - r.inferiortemporal, 8 - r.precentral.}
\label{fig:delta_1110_day_GC}
\end{center}
\end{figure}

Conditioning on the modal order of the process for both series in each subject, we obtain samples from the posterior distributions of the autoregressive coefficient matrices. The $(i,j)$-th element in the autoregressive matrix at lag-$s$, $\phi_{s,ij}$, governs the effect of the $j$-th variable at time $t-s$ on the $i$-th variable at time $t$. If $\phi_{s,ij}$ is non-zero we say that variable $j$ Granger-causes variable $i$ at lag $s$; this Granger-causal connection can be represented in a directed network, called a Granger causality plot, through an edge from vertex $j$ to vertex $i$. Conditional on the posterior modal order, $p^\ast=2$, Figures \ref{fig:beta_1110_day_GC} and \ref{fig:delta_1110_day_GC} show the Granger causality plots at lags 1 and 2 for individual A in the beta and delta bands, respectively. In these plots, an autoregressive coefficient is visualized as non-zero whenever zero lies outside the 50\% equi-tailed Bayesian credible interval; the thickness of the edges representing non-zero coefficients are representations of the absolute value of the posterior mean. The coordinates of the vertices, representing the different brain regions, correspond to the $x$ and $y$ coordinates of the centre of the region using the Desikan-Killiany atlas. A noticeable feature of these Granger causality plots is the higher number of connections in the delta band compared to the beta band. This was common across all individuals \citep[see Section S4.3 of the Supplementary Materials][]{supplMat}, and may indicate more localized processes underpinning the delta rhythms that interact with each other, whereas the beta rhythms in each region may be more driven by common processes. However future work has to confirm if this feature is a result of the epilepsy, or medication.




\subsection{Decomposition into latent series}

Using classic theory of time-series decompositions \citep[][]{prado_thesis_1998}, a VAR$_m$($p$) process can be decomposed into $pm$ latent series. These series correspond to the $pm$ distinct eigenvalues of the companion matrix which arises from the representation of the model as a VAR$_{mp}$($1$) process. Suppose there are $c$ complex conjugate pairs of eigenvalues denoted $r_j e^{\pm i \omega_j}$, $j=1,\ldots,c$, and $pm - 2c$ real eigenvalues denoted $r_j$, $j=2c+1, \ldots, pm$ where $r_j > 0$ and $\omega_j \in [0, \pi)$. The latent decomposition of $\vec{y}_t = (y_{t1}, \ldots, y_{tm})^\T$ then takes the form
\begin{equation*}
y_{ti} = \sum_{j=1}^c z_{tij} + \sum_{j=2c+1}^{pm} x_{tij}
\end{equation*}
where $z_{tij}$ and $x_{tij}$ are real-valued processes corresponding to the $j$th pair of complex eigenvalues and the $j$th real eigenvalue, respectively. The process $z_{tij}$ has similar behaviour to an AR(2) process with coefficients $2 r_j \cos \omega_j$ and $-r_j^2$ and is therefore quasi-periodic with characteristic frequency $\omega_j$ and modulus $r_j$. This holds for all dimensions $i=1,\ldots,m$, though the time-varying amplitude and phase are different for each $i$. Similarly, the process $x_{tij}$ follows an AR(1) structure with coefficient $r_j$ for all $i=1,\ldots,m$. The innovations that drive the $z_{tij}$ and $x_{tij}$ processes are correlated and arise from the error terms $\vec{\epsilon}_t$ in the original model.

The quasi-periodic series arising from the complex conjugate pairs of eigenvalues are of particular interest as they can capture the cyclical patterns that are key to understanding variation in brain activity. The pairs of complex eigenvalues, $r_j e^{\pm i \omega_j}$, $j=1,\ldots,c$, are not identifiable as the model remains unchanged under any permutation of their labelling. However, identification can be achieved by applying an ordering constraint, for example, based on the modulus or the argument. Imposing the constraint $\omega_1 < \omega_2 < \cdots < \omega_c$, the quasi-periodic series $z_{tij}$ are ordered by decreasing period $2 \pi / \omega_j$. 

\begin{figure}[!tb]
\begin{center}
\includegraphics[width=1\textwidth]{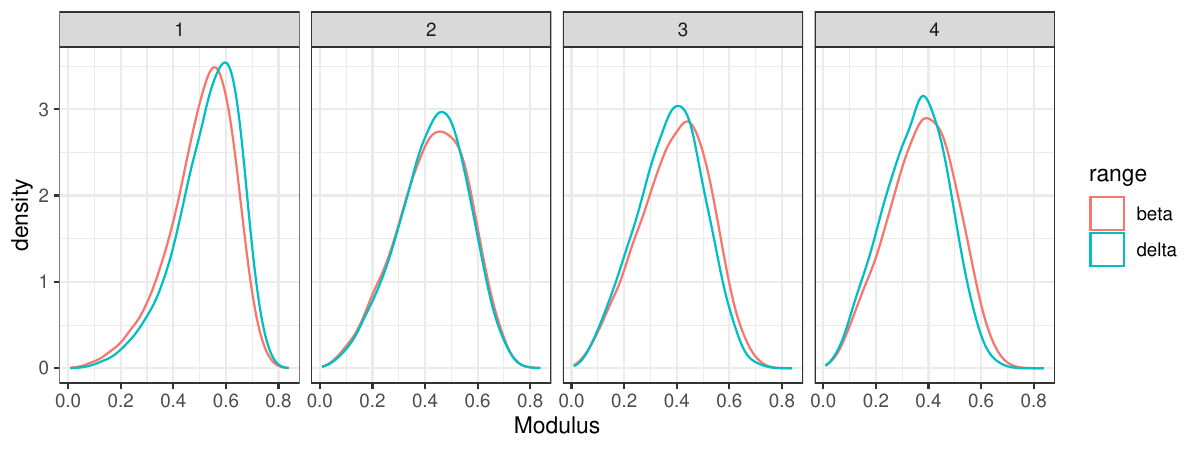}
\caption{Posterior densities for the moduli of the first four quasi-periodic series for individual A.}
\label{fig:1110_moduli_day}
\end{center}
\end{figure}

\begin{figure}[!tb]
\begin{center}
\includegraphics[width=1\textwidth]{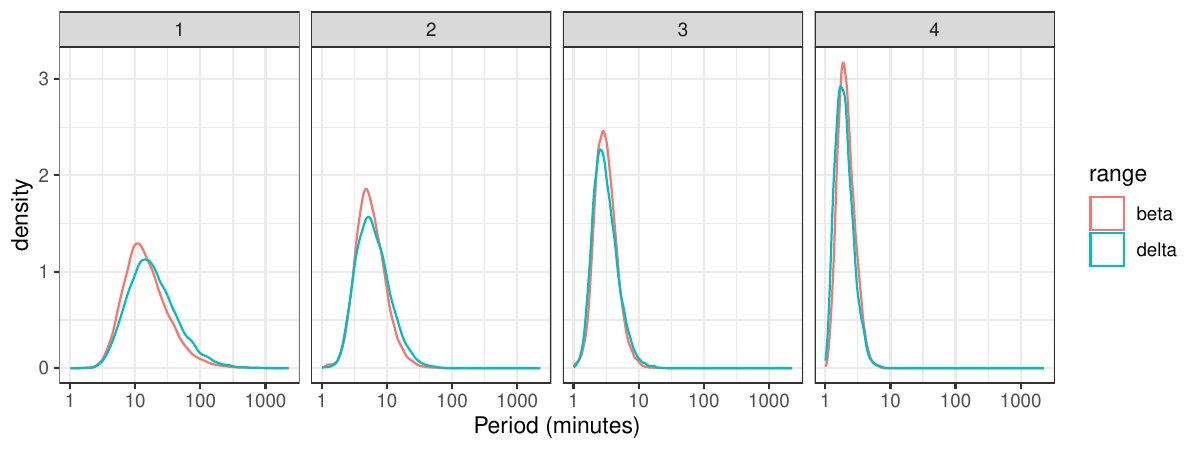}
\caption{Posterior densities for the periods of the first four quasi-periodic series for individual A.}
\label{fig:1110_periods_day}
\end{center}
\end{figure}

For individual A the posteriors for the moduli and periods of the first four quasi-periodic series are presented in Figures \ref{fig:1110_moduli_day} and \ref{fig:1110_periods_day}. We note that the $z_{tij}$ with highest period also have highest modulus and might therefore be regarded as the dominating latent series. Corresponding figures for the other individuals are provided in Section S4.4 of the Supplementary Materials \citep[][]{supplMat}. Across individuals, a common feature is that the posterior for the period of the dominating latent series in each band has its mean at around 20 minutes; for example, for individual A, the posterior means in the beta and delta bands are 22.83 and 29.51 minutes, with 95\% equi-tailed Bayesian credible intervals of $(4.04, 96.24)$ and $(4.32, 131.05)$ minutes, respectively. It is also noticeable that though there are some differences between the moduli of the series in the delta band compared to the beta band, there is very little difference between the corresponding periods. Again, this feature is replicated across all individuals. We elaborate further on this observation in Section \ref{sec:discussion}.


\section{\label{sec:discussion}Discussion}

We have proposed a hierarchical Bayesian model, with accompanying model-fitting methodology, which allows inference on the order of a stationary vector autoregression. This is based on an unconstrained reparameterization of the stationary model in terms of a set of transformed partial autocorrelation matrices \citep{heaps_enforcing_2022} whose properties can be exploited in the design of the prior. In particular, we capitalize on the nested structure of the new parameterization by constructing an overparameterized hierarchical model which shrinks unnecessary, high-order terms to zero; by identifying the lag beyond which the partial autocorrelation parameters become effectively equal to zero, we can then learn the order of the process. Further, using the relationship between the spectral norm of a partial autocorrelation matrix and its unconstrained counterpart, the prior is chosen to increasingly shrink the partial autocorrelation matrices at higher lags towards zero through a multiplicative gamma process for the unconstrained matrices. 

An efficient Hamiltonian Monte Carlo sampler for computational inference was proposed and implemented through Stan, with accompanying code to allow easy dissemination into other fields. The interpretability of the reparameterization allowed use of classical theory on the distribution of the estimators of the partial autocorrelation function to make a judgement about which sampled partial autocorrelation matrices are approximately equal to zero. An associated truncation criterion determines the number of non-zero partial autocorrelation matrices, allowing posterior inference on the order of the process in a manner which accounts for the scale, dimension and length of the time-series. This truncation criterion is independent of the choice of prior and could be applied in conjunction with any prior for an overparameterized stationary vector autoregression.

We applied our methodology to a series of simulation experiments in which data sets of various lengths $n$ were sampled from various stationary \VAR{m}{p} models. For all values of $m$, $p$ and $n$ considered, the posterior for the effective order of the process was highly concentrated around the known model order. We then applied our methodology to iEEG data from recordings at various locations in the brain. Conditioning on the posterior modal order of these processes allowed physiological insight in a number of directions. By constructing Granger causality plots, we were able to highlight relationships between activity in different regions of the brain. Similarly, by constructing the latent decomposition of the series, we were able to identify underlying quasi-periodic structure. In particular, we found that the dominant latent component had a period that was around 20 minutes across all individuals in both the beta and the delta bands. This is consistent with ultradian rhythms of around 20 minutes which have previously been observed \citep[][]{panagiotopoulou_fluctuations_2022}. The similarity in the periods across the beta and delta bands indicate that there is a global cycle in the band power pattern, rather than a local cycle within a specific band. The similarities between subjects are striking, particularly the period of 20 minutes, and warrant future investigations into the possible biological mechanisms and potentially endogenous drivers \citep[][]{goh_episodic_2019-1}. However, as we only considered four subjects in this work, a larger study would be needed to confirm any biological interpretations, with a larger number of patients, longer recordings and accounting for the potential pathology present in these subjects.

Although our focus has been learning the order of stationary vector autoregressions with independent, Gaussian errors, these assumptions about the error process could be relaxed in a number of ways. First, we could readily allow higher probabilities of extreme values by assuming the errors arose from a heavy tailed distribution with finite first and second moments, such as the multivariate-$t$ distribution with degree of freedom parameter greater than two. Second, it may also be possible to allow for heteroskedasticity by modelling the errors according to a multivariate generalized autoregressive conditional heteroskedasticity (GARCH) process, such as that described by \citet{LM03}, in which additional constraints on the parameters of the error process guarantee second order stationarity. In this instance, however, because the mapping from the original parameter space $(\matr{\Sigma}, \matr{\Phi})$ to $\{\matr{\Sigma}, (\matr{P}_1, \ldots, \matr{P}_p) \}$ depends on the conditional variance $\matr{\Sigma}$ and this conditional variance would change at every time point, further work would be needed to investigate an alternative choice of positive definite matrix to replace $\vec{\matr{\Sigma}}$, and to establish the interpretation of the resulting matrices $\matr{P}_1, \ldots, \matr{P}_p$. Finally, we could allow the errors to be autocorrelated by considering the more general class of stable and invertible vector autoregressive moving average (VARMA) models. A vector autoregressive moving average model of order ($p,q$) has the form 
\begin{equation*}
\vec{y}_t = \matr{\phi}_1 \vec{y}_{t-1} + \ldots + \matr{\phi}_p \vec{y}_{t-p} + \vec{\epsilon}_t + \matr{\psi}_1\vec{\epsilon}_{t-1} + \ldots + \matr{\psi}_q\vec{\epsilon}_{t-q}
\end{equation*}
where $\vec{\epsilon}_t\sim \norm(\vec{0},\Sigma)$ for $t=q+1, \ldots, n$, the parameters $\matr{\psi}_i \in M_{m\times m}(\mathbb{R})$, $i = 1, \ldots, q,$ are $m\times m$ moving average coefficient matrices and the parameters $\matr{\phi}_1, \ldots, \matr{\phi}_p$ are the autoregressive coefficient matrices which lie in $\mathcal{C}_{p,m}$ for a stable process. In backshift notation, this can be written as $\phi(B)\vec{y}_t = \psi(B)\vec{\epsilon}_t$ where $\psi(u) = (\matr{I}_m + \matr{\psi}_1 u + \ldots + \matr{\psi}_q u^q)$, $u \in \mathbb{C}$, is the characteristic moving average polynomial. The VARMA process is invertible if and only if all roots of the equation $\det\{\psi(u)\}=0$ lie outside the unit circle. The subset of $M_{m\times m}(\mathbb{R})$ in which the matrices $-\matr{\psi}_1, \ldots, -\matr{\psi}_q$ satisfy this condition is simply $\mathcal{C}_{q,m}$. To constrain the parameters to this region, we can therefore use the reparameterizations discussed in Section~\ref{subsec:reparam} to first map from the moving average coefficient matrices and variance matrix to a set of matrices $\matr{Q}_1, \ldots, \matr{Q}_q$ with singular values less than one, and then to map to a set of unconstrained square matrices; \citep[see][Supplementary Materials]{heaps_enforcing_2022}. To allow inference on $q$, as well as $p$, a second multiplicative gamma process prior could then be used to increasingly shrink the unconstrained matrices at higher lags towards zero, again using a truncation criterion to determine which of the matrices are effectively zero. However, since the matrices $\matr{Q}_1, \ldots, \matr{Q}_q$ do not have a clear interpretation, making a principled choice of truncation criterion for the reparameterized moving average coefficients would not be straightforward. Moreover, since VARMA models can be approximated by higher order vector autoregressions, it is likely that posteriors would be multimodal, demanding development of a Markov chain Monte Carlo sampler that can move efficiently between modes. We defer further consideration of this challenging topic to future work.

Our ideas could also be extended to learn the order of processes that are only piecewise stationary. Indeed, an obvious limitation in the application to iEEG data was the necessity to pick out contiguous segments of data where stationarity was a plausible assumption. However, as remarked in Section~\ref{sec:intro}, stationary autoregressions often serve as building blocks in the construction of more complex models. Motivated by applications involving iEEG data where subjects transition between states of wakefulness and sleep, or states of normal brain activity and seizure, we are currently exploring a hidden Markov model in which a (locally) stationary vector autoregression describes the within-state dynamics. Such a model would be ideally suited to a wide variety of time-series where there are occasional step-changes in a process which otherwise appears to be mean reverting. 

\section*{Acknowledgements}
This work was supported by the Engineering and Physical Sciences Research Council (EPSRC), Centre for Doctoral Training in Cloud Computing for Big Data (grant number EP/L015358/1).  This work was also supported by the EPSRC (grant number EP/N510129/1) via the the Alan Turing Institute project ``Streaming data modelling for real-time monitoring and forecasting''. The authors also gratefully acknowledge two anonymous referees, an Associate Editor and the Editor for their constructive comments that improved the quality of this paper.

\bigskip
\begin{center}
{\large\bf SUPPLEMENTARY MATERIAL}
\end{center}

\begin{description}
\sloppy
\item[Supporting information:] Further discussion of the stationary region, a numerical example of the reparameterization over the stationary region, additional details on the prior specification for the simulated and EEG applications, more extensive discussion of posterior sensitivity to the choice of truncation criterion, and additional plots from the EEG application. (.pdf file)

\item[Data, R code and Stan program:] The simulated data from Section~\ref{sec:simulation_studies}, R scripts for simulating data, running Stan and post-processing the Stan output, as well as the Stan program for fitting the model can be found at the GitHub repository \url{https://github.com/rachelbinks/Bayesian-VAR-order-determination}. (url)

\end{description}

\bibliographystyle{chicago}
\bibliography{references.bib}

\end{document}


\maketitle

\begin{abstract}
Supplementary material contained in this note includes further discussion of the stationary region, a numerical example of the reparameterization over the stationary region, additional details on the prior specification for the simulated and EEG applications, more extensive discussion of posterior sensitivity to the choice of truncation criterion, and additional plots from the EEG application.
\end{abstract}

\section{\label{sec:stat_region}Illustration of stationary region}
\begin{figure}[!th]
\begin{center}
\includegraphics[width=0.8\textwidth]{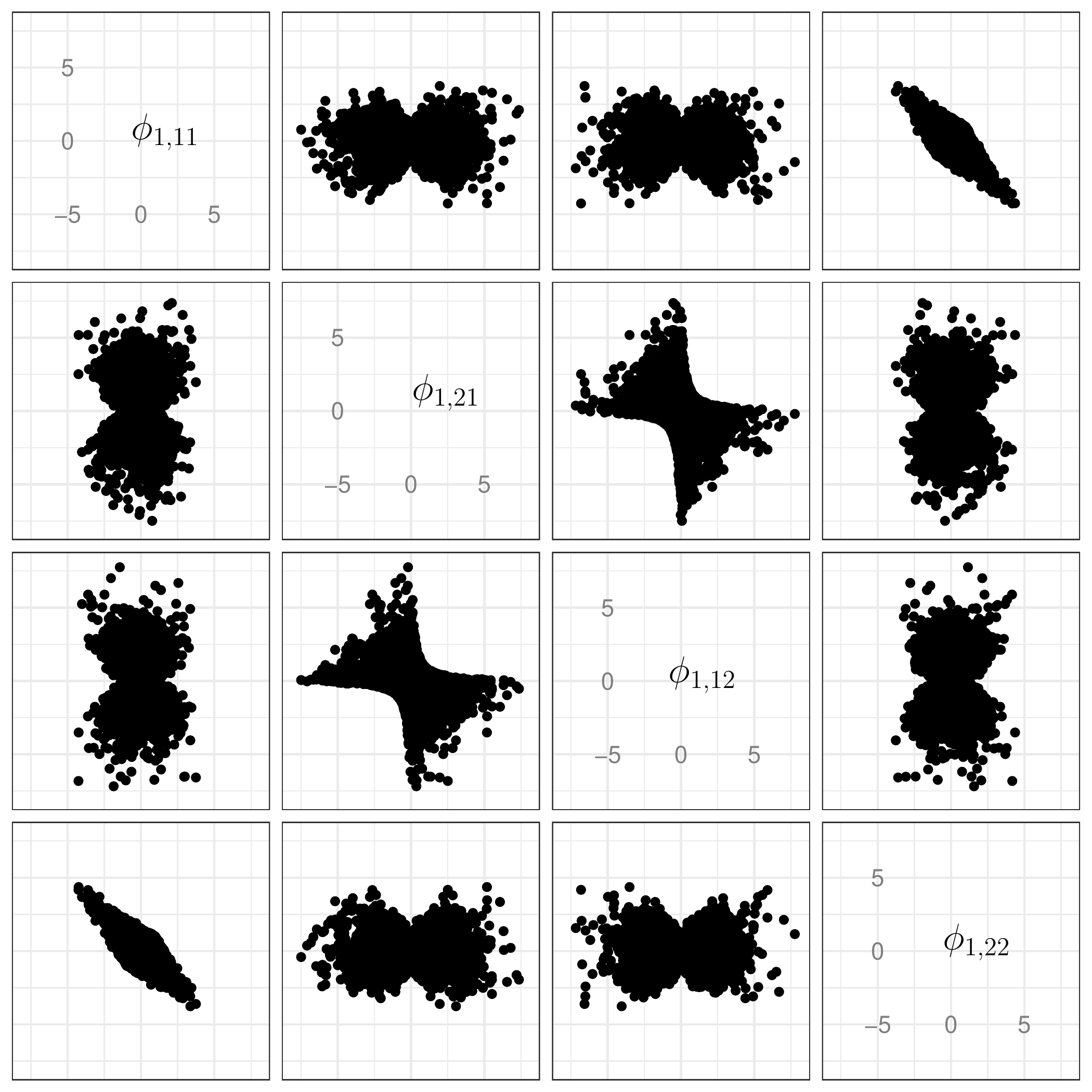}
\caption{Scatterplot matrix showing draws from a diffuse distribution over the stationary region for the elements of $\matr{\phi}_1$ in the \VAR{2}{1} process.}
\label{fig:stat_region}
\end{center}
\end{figure}

As discussed in Section~2 of the main manuscript, a vector autoregression is stable if and only if the roots of the equation $\det\{ \matr{\phi}(u) \} = 0$ lie outside the unit circle, where $\matr{\phi}(u) = (\matr{I}_m - \matr{\phi}_1 u - \ldots - \matr{\phi}_p u^p)$, $u \in \mathbb{C}$, is the characteristic polynomial. This condition constrains the parameter space to a region referred to as the stationary region, $\mathcal{C}_{p,m}$. When $m=1$ and $p=1$ or $p=2$, the stationary region is simple, with $\mathcal{C}_{1,1}$
 representing the interval $(-1, 1)$ and $\mathcal{C}_{2,1}$ a triangle in the $(\phi_1,\phi_2)$-plane. However, increasing $m$ or $p$ further increases the complexity of the polynomial equation $\det\{ \matr{\phi}(u) \} = 0$ and hence the geometry of the stationary region. This can be visualized even in the simplest vector autoregressive process where $m = 2$ and $p = 1$:
\begin{equation*}
\vec{y}_t = \matr{\phi}_1 \vec{y}_{t-1} + \vec{\epsilon}_t, \quad \vec{\epsilon}_t \sim \norm_2(\vecn{0}, \matr{\Sigma}).
\end{equation*}
In this case, the constraint $\matr{\Phi} \equiv \matr{\phi}_1 \in \mathcal{C}_{1,2}$ is equivalent to restricting the spectral radius of $\matr{\phi}_1$, $\rho(\matr{\phi}_1)$, to be less than one, which is visualized in Figure~\ref{fig:stat_region}.

\section{\label{sec:toy_example_reparameterization}Example of reparameterization over the stationary region}
As discussed in Section~2.2 of the main manuscript, due to the complex geometry of the stationary region, we follow \citet{heaps_enforcing_2022} in reparameterizing the model over the stationary region in terms of a set of interpretable, unconstrained parameters. In this section, we provide a toy example of how this mapping works, following the steps of the reparameterization which proceeds via recursion. In the example we outline each step for the simple case where $m=2$ and $p=2$. Throughout the example, the presented numerical values are rounded to two decimal places, but the unrounded values are used for all calculations. Let $\vec{y}_t$ follow a stationary VAR$_2$($2)$ process such that
\begin{equation*}
\vec{y}_t = \matr{\phi}_1 \vec{y}_{t-1} + \matr{\phi}_2\vec{y}_{t-2} + \vec{\epsilon}_t, \quad \vec{\epsilon}_t \sim \norm_2(\vecn{0}, \matr{\Sigma}),
\end{equation*}
with
\begin{equation*} 
\matr{\phi_1} = \begin{pmatrix} 0.1 & 0.1 \\ 0.2 & 0.2 \end{pmatrix}, \hspace{0.5cm} \matr{\phi_2} = \begin{pmatrix} 0.3 & 0.2 \\ 0.1 & 0.2 \end{pmatrix} \hspace{0.5cm} \textrm{and} \hspace{0.5cm} \matr{\Sigma} = \begin{pmatrix} 1& 0 \\ 0 & 1 \end{pmatrix}.
\end{equation*}

\subsection{\label{subsec:forward_mapping_example}Forward mapping}

In generality, the forward mapping algorithm is as follows:
\begin{enumerate}
		\item For $i=0, \ldots, p$, compute the autocovariances $\matr{\Gamma}_i = \Cov(\vec{y}_t, \vec{y}_{t+i})$ from $(\matr{\Sigma}, \matr{\Phi})$.  $\matr{\Gamma}_0, \ldots, \matr{\Gamma}_{p-1}$ can be found by representing the \VAR{m}{p} process as a \VAR{mp}{1} process and computing its stationary variance. This results in a discrete Lyapunov equation which can be solved using vectorization and Kronecker product operators. Then, $\matr{\Gamma}_p$ can be found using the Yule-Walker equations for the order $p$ process. Further details on such calculations can be found in Chapter~2 of \citet{Lut06}.
		
		\item Compute the partial autocorrelation matrices $(\matr{P}_1, \ldots, \matr{P}_p)$ from $\matr{\Phi}$ and $(\matr{\Gamma}_0, \ldots, \matr{\Gamma}_p)$:
		\begin{enumerate}
			\item Initialize: construct $\matr{\Sigma}_0 = \matr{\Sigma}_0^* = \matr{\Gamma}_0$ then calculate the symmetric matrix-square-root factorization such that 
			$$\matr{\Sigma}_0 = \matr{\Sigma}_0^{\frac{1}{2}}\matr{\Sigma}_0^{\frac{1}{2}} = \matr{\Sigma}_0^{*\frac{1}{2}}\matr{\Sigma}_0^{*\frac{1}{2}} = \matr{\Sigma}_0^*.$$
			\item Recursion: for each $s=0,\ldots,p-1$:
			\begin{enumerate}
				\item Compute $\matr{\phi}_{s+1,s+1}$ using
				\begin{equation*}  \matr{\phi}_{s+1,s+1}=(\matr{\Gamma}_{s+1}^{\T} - \matr{\phi}_{s1} \matr{\Gamma}_s^{\T} - \cdots - \matr{\phi}_{ss}\matr{\Gamma}_1^{\T})\matr{\Sigma}_s^{*-1}
				\end{equation*}
				and $\matr{\phi}_{s+1, s+1}^*$ using
				\begin{equation*}  
					\matr{\phi}_{s+1,s+1}^*=(\matr{\Gamma}_{s+1} - \matr{\phi}_{s1}^* \matr{\Gamma}_s - \cdots - \matr{\phi}_{ss}^*\matr{\Gamma}_1)\matr{\Sigma}_s^{-1}.
				\end{equation*}
				In the case where $s=0$ this simplifies to give $$\matr{\phi}_{11} = \matr{\Gamma}_1^{\T}\matr{\Sigma}_0^{*-1} = \matr{\Gamma}_1^{\T}\matr{\Gamma}_0^{-1} \quad \text{and} \quad \matr{\phi}_{11}^* = \matr{\Gamma}_1\matr{\Sigma}_0^{-1} = \matr{\Gamma}_1\matr{\Gamma}_0^{-1}.$$
				
				\item If $s>0$, for $i=1, \ldots, s$, compute $\matr{\phi}_{s+1,i}$ using  
				\begin{equation*}  \matr{\phi}_{s+1,i}=\matr{\phi}_{si}-\matr{\phi}_{s+1,s+1}\matr{\phi}_{s,s-i+1}^*
				\end{equation*} 
				and $\matr{\phi}_{s+1,i}^*$ using
				$$\matr{\phi}_{s+1,i}^* = \matr{\phi}_{si}^* - \matr{\phi}_{s+1, s+1}^*\matr{\phi}_{s,s-i+1}.$$			
				
				\item Compute the $(s+1)$th partial autocorrelation $\matr{P}_{s+1}$ using either
				\begin{equation*}
					 \matr{P}_{s+1}=\matr{\Sigma}_s^{-\frac{1}{2}}\matr{\phi}_{s+1,s+1}\matr{\Sigma}_s^{*\frac{1}{2}}\quad \text{or} \quad 
				\matr{P}_{s+1}=(\matr{\Sigma}_s^{*-\frac{1}{2}}\matr{\phi}_{s+1,s+1}^*\matr{\Sigma}_s^{\frac{1}{2}})^{\T}.
				\end{equation*} 
				\item If $s<p-1,$ compute $\matr{\Sigma}_{s+1}$ using
				$$\matr{\Sigma}_{s+1} = \matr{\Gamma}_0 - \matr{\phi}_{s+1,1}\matr{\Gamma}_1 - \cdots- \matr{\phi}_{s+1, s+1}\matr{\Gamma}_{s+1}$$
				and $\matr{\Sigma}_{s+1}^*$ using
				$$\matr{\Sigma}_{s+1}^* = \matr{\Gamma}_0 - \matr{\phi}_{s+1,1}^*\matr{\Gamma}_1^{\T} - \cdots - \matr{\phi}_{s+1, s+1}^*\matr{\Gamma}_{s+1}^{\T}.$$
				Then calculate the symmetric matrix-square-roots such that 
				$$\matr{\Sigma}_{s+1} = \matr{\Sigma}_{s+1}^{\frac{1}{2}}\matr{\Sigma}_{s+1}^{\frac{1}{2}} \quad \text{and} \quad \matr{\Sigma}_{s+1}^* = \matr{\Sigma}_{s+1}^{*\frac{1}{2}}\matr{\Sigma}_{s+1}^{*\frac{1}{2}}.$$
			\end{enumerate}
		\end{enumerate}
	\end{enumerate}

In our example, this works as follows:
\begin{enumerate}
\item By solving the system
\begin{equation*}
\left\{ 
\matr{I}_{16} -
\begin{pmatrix}
\matr{\phi}_1 &\matr{\phi}_2\\
\matr{I}_2    &\matrn{0}_2
\end{pmatrix} \otimes
\begin{pmatrix}
\matr{\phi}_1 &\matr{\phi}_2\\
\matr{I}_2    &\matrn{0}_2
\end{pmatrix}
\right\}^{-1} \vect
\begin{pmatrix}
\matr{\Sigma} &\matrn{0}_2\\
\matrn{0}_2   &\matrn{0}_2
\end{pmatrix},
\end{equation*}
reshaping the vector into a $4 \times 4$ matrix and taking $\matr{\Gamma}_0$ and $\matr{\Gamma}_1$ as the $(2 \times 2)$ matrices in the top and bottom blocks on the left, we obtain
\begin{equation*} 
\matr{\Gamma}_0 = \begin{pmatrix} 1.31 & 0.28 \\ 0.28 & 1.30 \end{pmatrix} \hspace{0.5cm} \textrm{and} \hspace{0.5cm} \matr{\Gamma}_1 = \begin{pmatrix} 0.34 & 0.43 \\ 0.38 & 0.45 \end{pmatrix}.
\end{equation*}
Then $\matr{\Gamma}_2$ is obtained through
\begin{equation*}
\matr{\Gamma}_2 = \matr{\Gamma}_1 \matr{\phi}_1^\T + \matr{\Gamma}_0 \matr{\phi}_2^\T = \begin{pmatrix} 0.53& 0.34 \\ 0.43 & 0.45 \end{pmatrix}.
\end{equation*}
		\item  \begin{enumerate}
			\item
			\begin{equation*} \matr{\Sigma}_0 = \matr{\Sigma}_0^* = \matr{\Gamma}_0 = \begin{pmatrix} 1.31 & 0.28 \\ 0.28 & 1.30 \end{pmatrix}. \end{equation*}

			The symmetric matrix-square-root factorizations can be calculated as 
 			\begin{equation*} \matr{\Sigma}_0^{1/2} = \matr{\Sigma}_0^{* \, 1/2} = \begin{pmatrix} 1.14 & 0.12 \\ 0.12 & 1.13 \end{pmatrix}, \end{equation*}
			using, for example, the \texttt{sqrtm} function from the \texttt{expm} package in R. 

			\item 
			\begin{enumerate}
				\item
				Take $s=0$. 
				\begin{equation*} \matr{\phi}_{11} = \matr{\Gamma}_1^{\T}\matr{\Sigma}_0^{*-1} = \begin{pmatrix} 0.20 & 0.25 \\ 0.26 & 0.29 \end{pmatrix}, \end{equation*}

				\begin{equation*} \matr{\phi}_{11}^* = \matr{\Gamma}_1\matr{\Gamma}_0^{-1} = \begin{pmatrix} 0.19 & 0.29 \\ 0.22 & 0.30 \end{pmatrix}. \end{equation*}
 				
				\item As $s=0$, this step is skipped.

				\item 
				\begin{equation*} \matr{P}_1 = \matr{\Sigma}_0^{-1/2}\matr{\phi}_{11}\matr{\Sigma}_0^{* \, 1/2} = \begin{pmatrix} 0.20 & 0.23 \\ 0.28 & 0.29 \end{pmatrix}. \end{equation*}

				\item \begin{equation*} \matr{\Sigma}_1 = \matr{\Gamma}_0 - \matr{\phi}_{11}\matr{\Gamma}_1 = \begin{pmatrix} 1.15 & 0.08 \\ 0.08 & 1.06 \end{pmatrix}, \end{equation*}
				\begin{equation*} \matr{\Sigma}_1^* = \matr{\Gamma}_0 - \matr{\phi}_{11}^* \matr{\Gamma}_1^{\T} = \begin{pmatrix} 1.13 & 0.08 \\ 0.08 & 1.08 \end{pmatrix}, \end{equation*}
				\begin{equation*} \matr{\Sigma}_1^{1/2} = \begin{pmatrix} 1.07 & 0.04\\ 0.04 & 1.03 \end{pmatrix}, \end{equation*}
				\begin{equation*} \matr{\Sigma}_1^{* \, 1/2} = \begin{pmatrix} 1.06 & 0.04 \\ 0.04 & 1.04 \end{pmatrix}. \end{equation*}
			\end{enumerate}
			\begin{enumerate}
				\item Take $s=1$. 
				\begin{equation*} \matr{\phi}_{22} = (\matr{\Gamma}_2^{\T}- \matr{\phi}_{11}\matr{\Gamma}_1^{\T})\matr{\Sigma}_1^{*-1} = \begin{pmatrix} 0.30 & 0.20 \\ 0.10 & 0.20 \end{pmatrix}, \end{equation*}
				\begin{equation*} \matr{\phi}_{22}^* =(\matr{\Gamma}_2- \matr{\phi}_{11}^*\matr{\Gamma}_1)\matr{\Sigma}_1^{-1} = \begin{pmatrix} 0.30 & 0.10 \\ 0.19 & 0.19 \end{pmatrix}. \end{equation*}

				\item \begin{equation*} \matr{\phi}_{21} = \matr{\phi}_{11} - \matr{\phi}_{22}\matr{\phi}_{11}^* = \begin{pmatrix} 0.10 & 0.10 \\0.20 & 0.20 \end{pmatrix}, \end{equation*}
				          \begin{equation*} \matr{\phi}_{21}^* = \matr{\phi}_{11}^* - \matr{\phi}_{22}^*\matr{\phi}_{11} = \begin{pmatrix} 0.11 & 0.19 \\ 0.13 & 0.19 \end{pmatrix}. \end{equation*}
	
				\item \begin{equation*} \matr{P}_2 = \matr{\Sigma}_1^{-1/2}\matr{\phi}_{22}\matr{\Sigma}_1^{* \, 1/2} = \begin{pmatrix} 0.30 & 0.20 \\ 0.10 & 0.20 \end{pmatrix}. \end{equation*}
				\item As $s = p - 1$, this step is skipped.
			\end{enumerate}
		\end{enumerate}
\end{enumerate}
The final (rounded) partial autocorrelation matrices are therefore 
\begin{equation*} \matr{P}_1 = \begin{pmatrix} 0.20 & 0.23 \\ 0.28 & 0.29 \end{pmatrix} \hspace{0.5cm} \textrm{and} \hspace{0.5cm} \matr{P}_2 = \begin{pmatrix} 0.30 & 0.20 \\ 0.10 & 0.20 \end{pmatrix}, \end{equation*} which both have singular values less than one, satisfying the stationarity condition under this parameterization (by construction). 

As discussed in the main manuscript, a second transformation then maps each of the partial autocorrelation matrices to an unconstrained matrix through 
\begin{equation*} A = (I_m - \matr{P}\matr{P}^{\T})^{-1/2}\matr{P}. \end{equation*}
In our example, this results in 
\begin{equation*} 
A_1 = (I_2 - \matr{P}_1\matr{P}_1^{\T})^{-1/2}\matr{P}_1 = \begin{pmatrix} 0.23 & 0.27 \\ 0.32 & 0.34 \end{pmatrix} \end{equation*}
and 
\begin{equation*}A_2 = (I_2 - \matr{P}_2\matr{P}_2^{\T})^{-1/2}\matr{P}_2 = \begin{pmatrix} 0.33 & 0.22 \\ 0.11 & 0.21 \end{pmatrix}. \end{equation*}
\subsection{\label{subsec:backward_mapping_example}Backward mapping}
The final mapping in the previous subsection can be reversed using the mapping $$\matr{P} = (I_m + AA^{\T})^{-\frac{1}{2}}A.$$ 

Starting with the (unrounded) unconstrained matrices $A_1$ and $A_2$ calculated in the previous subsection, in our case this mapping returns the values of $\matr{P}_1$ and $\matr{P}_2$ calculated in the previous subsection:
$$\matr{P}_1 = (I_m + A_1A_1^{\T})^{-\frac{1}{2}}A_1 = \begin{pmatrix} 0.20 & 0.23 \\ 0.28 & 0.29 \end{pmatrix}, $$
$$\matr{P}_2 = (I_m + A_2A_2^{\T})^{-\frac{1}{2}}A_2 = \begin{pmatrix} 0.30 & 0.20 \\ 0.10 & 0.20 \end{pmatrix}. $$

As discussed in \citet{heaps_enforcing_2022}, the backward mapping algorithm from the partial autocorrelation matrices then proceeds, in generality, as follows:

\begin{enumerate}
		\item Compute the stationary variance $\matr{\Gamma}_0$ from $\{\matr{\Sigma}, (\matr{P}_1, \ldots, \matr{P}_p) \}$:
		\begin{enumerate}
			\item Initialize: let $\matr{\Sigma}_p=\matr{\Sigma}$ with the corresponding symmetric matrix-square-root factorization being such that $\matr{\Sigma}_p = \matr{\Sigma}_p^{\frac{1}{2}}\matr{\Sigma}_p^{\frac{1}{2}}$.
			\item Recursion: for $s=p-1,\ldots,0$, construct the symmetric matrix $\matr{\Sigma}_s^{\frac{1}{2}}$ according to
			\begin{equation*}
				\matr{\Sigma}_s^{\frac{1}{2}} = \matr{B}_{s+1}(\matr{B}_{s+1}^{-1} \matr{\Sigma}_{s+1} \matr{B}_{s+1}^{-1})^{1/2} \matr{B}_{s+1}
			\end{equation*}
		
			where $\matr{B}_{s+1}=(\matr{I}_m - \matr{P}_{s+1}\matr{P}_{s+1}^{\T})^{-1/2}$ then compute $\matr{\Sigma_s} = \matr{\Sigma}_s^{\frac{1}{2}}\matr{\Sigma}_s^{\frac{1}{2}}.$
			\item Output: set $\matr{\Gamma}_0 = \matr{\Sigma}_0$.
		\end{enumerate}
		\item Compute the autoregressive coefficient matrices in $\matr{\Phi}$ from $(\matr{P}_1, \ldots, \matr{P}_p)$ and $\matr{\Gamma}_0$:
		\begin{enumerate}
			\item Initialize: set $\matr{\Sigma}_0= \matr{\Sigma}_0^* =\matr{\Gamma}_0$ with corresponding symmetric matrix-square-root factorization such that $\matr{\Sigma}_0 = \matr{\Sigma}_0^{\frac{1}{2}} \matr{\Sigma}_0^{\frac{1}{2}} = \matr{\Sigma}_0^{*\frac{1}{2}}\matr{\Sigma}_0^{*\frac{1}{2}} = \matr{\Sigma}_0^*.$
			\item  Recursion: for each $s=0,\ldots,p-1$:
			\begin{enumerate}
				\item Compute $\matr{\phi}_{s+1,s+1}$ using
				\begin{equation*}
					\matr{\phi}_{s+1,s+1}=\matr{\Sigma}_s^{\frac{1}{2}}\matr{P}_{s+1}\matr{\Sigma}_s^{*-\frac{1}{2}}
				\end{equation*}
				and $\matr{\phi}_{s+1, s+1}^*$ using
				$$\matr{\phi}_{s+1, s+1}^* = \matr{\Sigma}_s^{*\frac{1}{2}}\matr{P}_{s+1}^{\T}\matr{\Sigma}_s^{-\frac{1}{2}}.$$
				
				\item If $s>0$, for $i=1, \ldots, s$ compute $\matr{\phi}_{s+1,i}$ using 
				\begin{equation*} 
					\matr{\phi}_{s+1,i}=\matr{\phi}_{si}-\matr{\phi}_{s+1,s+1}\matr{\phi}_{s,s-i+1}^*
				\end{equation*}
				and $\matr{\phi}_{s+1,i}^*$ using
				$$\matr{\phi}_{s+1,i}^* = \matr{\phi}_{si}^* - \matr{\phi}_{s+1, s+1}^*\matr{\phi}_{s,s-i+1}.$$
				\item Compute $\matr{\Sigma}_{s+1}$ using 
				\begin{equation*}
					\matr{\Sigma}_{s+1} = \matr{\Sigma}_s - \matr{\phi}_{s+1,s+1}\matr{\Sigma}_s^*\matr{\phi}_{s+1, s+1}^{\T}
				\end{equation*}
				and $\matr{\Sigma}_{s+1}^*$ using
				$$\matr{\Sigma}_{s+1}^* = \matr{\Sigma}_s^* - \matr{\phi}_{s+1, s+1}^*\matr{\Sigma}_s\matr{\phi}_{s+1, s+1}^{*\T}.$$
				Then calculate the symmetric matrix-square-roots such that
				$$\matr{\Sigma}_{s+1} = \matr{\Sigma}_{s+1}^{\frac{1}{2}}\matr{\Sigma}_{s+1}^{\frac{1}{2}} \quad \text{and} \quad \matr{\Sigma}_{s+1}^* = \matr{\Sigma}_{s+1}^{*\frac{1}{2}}\matr{\Sigma}_{s+1}^{*\frac{1}{2}}.$$
				\item Compute $\matr{\Gamma}_{s+1}$ using 
				\begin{equation*}
					\label{eq:gamma_s+1}
					\matr{\Gamma}_{s+1}^{\T}=\matr{\phi}_{s+1,s+1}\matr{\Sigma}_s^*+\matr{\phi}_{s1}\matr{\Gamma}_s^{\T}+\cdots+\matr{\phi}_{ss}\matr{\Gamma}_1^{\T}. 
				\end{equation*}
			\end{enumerate}
			\item Output: take $\matr{\phi}_i = \matr{\phi}_{pi}$ for $i=1, \ldots, p$ and from step 1(a) we have $\matr{\Sigma} = \matr{\Sigma}_p$.
		\end{enumerate}
		
	\end{enumerate}

In our example, using the (unrounded) values of $\matr{P}_1$ and $\matr{P}_2$, this works as follows:
\begin{enumerate}
\item
\begin{enumerate}
\item \begin{equation*} \matr{\Sigma}_2 =  \matr{\Sigma} = \begin{pmatrix} 1 & 0 \\ 0 & 1 \end{pmatrix}, \end{equation*}
\begin{equation*} \matr{\Sigma}^{1/2} = \begin{pmatrix} 1& 0 \\ 0 & 1 \end{pmatrix}. \end{equation*}
\item Take $ s= 1$.  Using
\begin{equation*} \matr{\Sigma}_1^{\frac{1}{2}} = \matr{B}_{2}(\matr{B}_{2}^{-1} \matr{\Sigma}_{2} \matr{B}_{2}^{-1})^{1/2} \matr{B}_{2}, \end{equation*}
where $\matr{B}_{2}=(\matr{I}_2 - \matr{P}_2 \matr{P}_2^{\T})^{-1/2}$ gives 
\begin{equation*} \matr{\Sigma}_1^{1/2} = \begin{pmatrix} 1.07 & 0.04\\ 0.04 & 1.03 \end{pmatrix}, \end{equation*}
\begin{equation*} \matr{\Sigma}_1 = \matr{\Sigma}_1^{1/2}\matr{\Sigma}_1^{1/2} = \begin{pmatrix} 1.15 & 0.08 \\ 0.08 & 1.06 \end{pmatrix}. \end{equation*}
Take $s=0$. Using 
\begin{equation*} \matr{\Sigma}_0^{\frac{1}{2}} = \matr{B}_{1}(\matr{B}_{1}^{-1} \matr{\Sigma}_{1} \matr{B}_{1}^{-1})^{1/2} \matr{B}_{1}, \end{equation*}
where $\matr{B}_{1}=(\matr{I}_2 - \matr{P}_1 \matr{P}_1^{\T})^{-1/2}$ gives 
\begin{equation*} \matr{\Sigma}_0^{1/2} = \begin{pmatrix} 1.14 & 0.12 \\ 0.12 & 1.13 \end{pmatrix}, \end{equation*}
\begin{equation*} \matr{\Sigma}_0 = \matr{\Sigma}_0^{1/2}\matr{\Sigma}_0^{1/2} = \begin{pmatrix} 1.31 & 0.28 \\ 0.28 & 1.30 \end{pmatrix}. \end{equation*}
\item \begin{equation*} \matr{\Gamma}_0 = \matr{\Sigma}_0 = \begin{pmatrix} 1.31 & 0.28 \\ 0.28 & 1.30 \end{pmatrix}. \end{equation*}
\end{enumerate}
\item 
\begin{enumerate}
\item
\begin{equation*} \matr{\Sigma}_0 = \matr{\Sigma}_0^* = \matr{\Gamma}_0 = \begin{pmatrix} 1.31 & 0.28 \\ 0.28 & 1.30 \end{pmatrix}, \end{equation*}
\begin{equation*} \matr{\Sigma}_0^{1/2} = \matr{\Sigma}_0^{* \, 1/2} = \begin{pmatrix} 1.14 & 0.12 \\ 0.12 & 1.13 \end{pmatrix}. \end{equation*}
\item Take $s=0$.
\begin{enumerate}
\item \begin{equation*} \matr{\phi}_{11} = \matr{\Sigma}_0^{1/2}\matr{P}_1\matr{\Sigma}_0^{* \, -1/2} = \begin{pmatrix} 0.20 & 0.25 \\ 0.26 & 0.29 \end{pmatrix}, \end{equation*}
\begin{equation*} \matr{\phi}_{11}^* = \matr{\Sigma}_0^{* \, 1/2} \matr{P}_1^{\T} \matr{\Sigma}_0^{-1/2} = \begin{pmatrix} 0.20 & 0.29 \\ 0.22 & 0.30 \end{pmatrix}. \end{equation*}

\item As $s=0$, this step is skipped.
\item \begin{equation*} \matr{\Sigma}_1 = \matr{\Sigma}_0 - \matr{\phi}_{11}\matr{\Sigma}_0^*\matr{\phi}_{11}^{\T} = \begin{pmatrix} 1.15 & 0.08 \\ 0.08 & 1.06 \end{pmatrix} ,\end{equation*}
\begin{equation*} \matr{\Sigma}_1^* = \matr{\Sigma}_0^* - \matr{\phi}_{11}^*\matr{\Sigma}_0\matr{\phi}_{11}^{*^{\T}} = \begin{pmatrix} 1.13 & 0.08 \\ 0.08 & 1.08 \end{pmatrix}, \end{equation*}
\begin{equation*} \matr{\Sigma}_1^{1/2} = \begin{pmatrix} 1.07 & 0.04 \\ 0.04 & 1.03 \end{pmatrix}, \end{equation*}
\begin{equation*} \matr{\Sigma}_1^{* \, 1/2} = \begin{pmatrix} 1.06 & 0.04 \\ 0.04 & 1.04 \end{pmatrix}. \end{equation*}
\item \begin{equation*} \matr{\Gamma}_1 = (\matr{\phi}_{11}\matr{\Sigma}_0^*)^{\T} = \begin{pmatrix} 0.34 & 0.43 \\ 0.38 & 0.45 \end{pmatrix}.\end{equation*}
\end{enumerate}
 Take $s=1$
\begin{enumerate}
\item \begin{equation*} \matr{\phi}_{22} = \matr{\Sigma}_1^{1/2} \matr{P}_2 \matr{\Sigma}_1^{* \, -1/2} = \begin{pmatrix} 0.30 & 0.20 \\ 0.10 & 0.20 \end{pmatrix}, \end{equation*}
\begin{equation*} \matr{\phi}_{22}^* = \matr{\Sigma}_1^{* \, 1/2} \matr{P}_2^{\T} \matr{\Sigma}_1^{-1/2} = \begin{pmatrix} 0.30 & 0.10 \\ 0.19 & 0.19 \end{pmatrix}.\end{equation*}
\item \begin{equation*} \matr{\phi}_{21} = \matr{\phi}_{11} - \matr{\phi}_{22}\matr{\phi}_{11}^* = \begin{pmatrix} 0.10 & 0.10 \\ 0.20 & 0.20 \end{pmatrix}, \end{equation*}
\begin{equation*} \matr{\phi}_{21}^* = \matr{\phi}_{11}^* - \matr{\phi}_{22}^*\matr{\phi}_{11} = \begin{pmatrix} 0.11 & 0.19 \\ 0.13 & 0.19 \end{pmatrix}. \end{equation*}
\item \begin{equation*} \matr{\Sigma}_2 = \matr{\Sigma}_1 - \matr{\phi}_{22}\matr{\Sigma}_1^* \matr{\phi}_{22}^{\T} = \begin{pmatrix} 1.00 & 0.00 \\ 0.00 & 1.00 \end{pmatrix}, \end{equation*}
\begin{equation*} \matr{\Sigma}_2^* = \matr{\Sigma}_1^* - \matr{\phi}_{22}^*\matr{\Sigma}_1\matr{\phi}_{22}^{*^{\T}} = \begin{pmatrix}1.01 & -0.02 \\ -0.02 & 0.99 \end{pmatrix}. \end{equation*}
\item \begin{equation*} \matr{\Gamma}_2 = (\matr{\phi}_{22}\matr{\Sigma}_1^* + \matr{\phi}_{11}\matr{\Gamma}_1^{\T})^{\T} = \begin{pmatrix} 0.53 & 0.34 \\ 0.43 & 0.45 \end{pmatrix}. \end{equation*}
\end{enumerate}
\item \begin{equation*} \matr{\phi}_1 = \matr{\phi}_{21} = \begin{pmatrix} 0.1 & 0.1 \\ 0.2 & 0.2 \end{pmatrix}, \end{equation*}
\begin{equation*} \matr{\phi}_2 = \matr{\phi}_{22} = \begin{pmatrix} 0.3 & 0.2 \\ 0.1 & 0.1 \end{pmatrix}, \end{equation*}
\begin{equation*} \matr{\Sigma} = \matr{\Sigma}_2 = \begin{pmatrix} 1 & 0 \\ 0 & 1 \end{pmatrix}. \end{equation*}
\end{enumerate}
\end{enumerate}
The final calculated parameters are the original parameters specified at the start of the previous subsection.

\section{Simulation experiments}

\subsection{\label{subsec:suppl_simulation_hyperparam}Choice of hyperparameters in the prior}
In the simulation experiments, we followed guidelines provided in \citet{durante_note_2017} to choose the hyperparameters of the multiplicative gamma process prior, taking $a_1 = 2.5$ and $a_2 = 3$. Additionally, we set $a=6$. These hyperparameter choices result in a prior for the elements of the unconstrained $\matr{A}_s$ matrices which have marginal variances less than one. As discussed in the Supplementary Materials of \citet{heaps_enforcing_2022}, this is necessary to avoid multimodality in the prior induced for the partial autocorrelation matrices. In the inverse Wishart distribution used as a prior for $\matr{\Sigma}$, the scale matrix is taken as $\matr{I}_m$ and the degrees of freedom as $m+4$, which ensures the variance is finite. 

\subsection{\label{subsec:suppl_sensitivity_beta}Sensitivity to choice of truncation criterion}
\begin{figure}[!th]
\begin{center}
\subfloat[][]{\includegraphics[width=.725\textwidth]{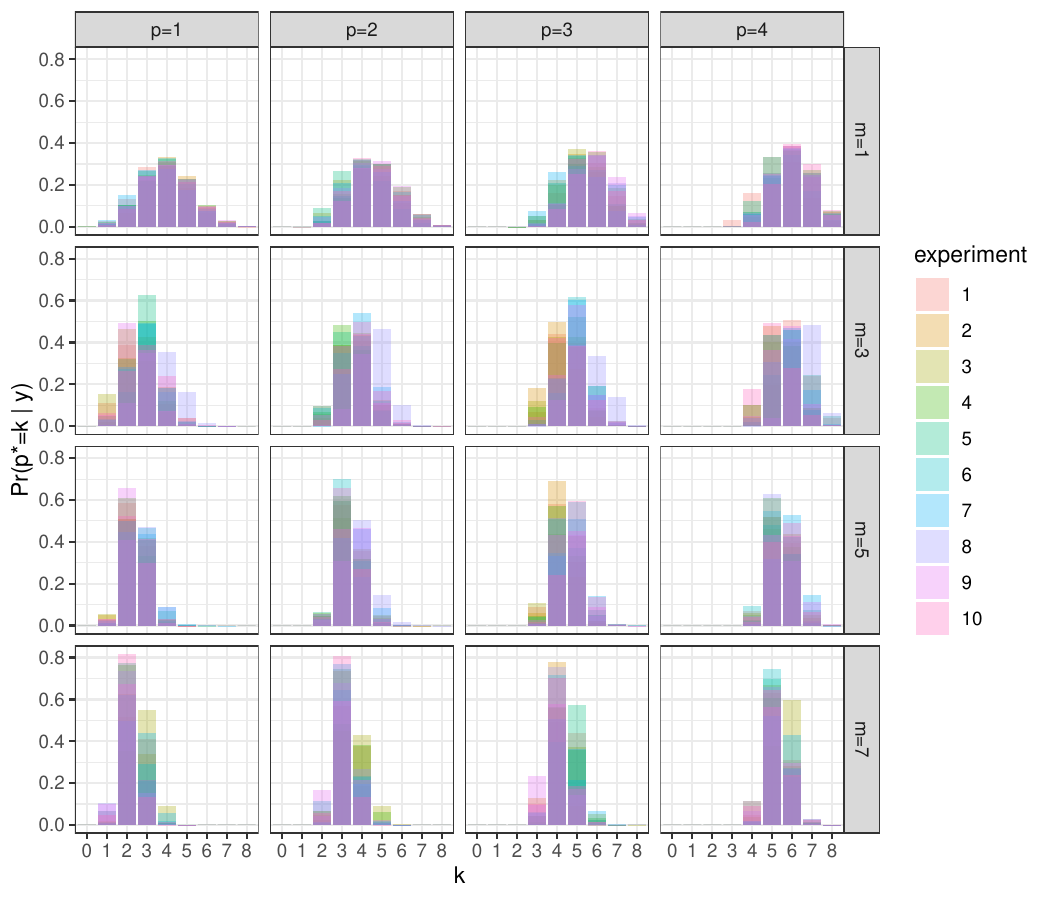}}\\
\subfloat[][]{\includegraphics[width=.725\textwidth]{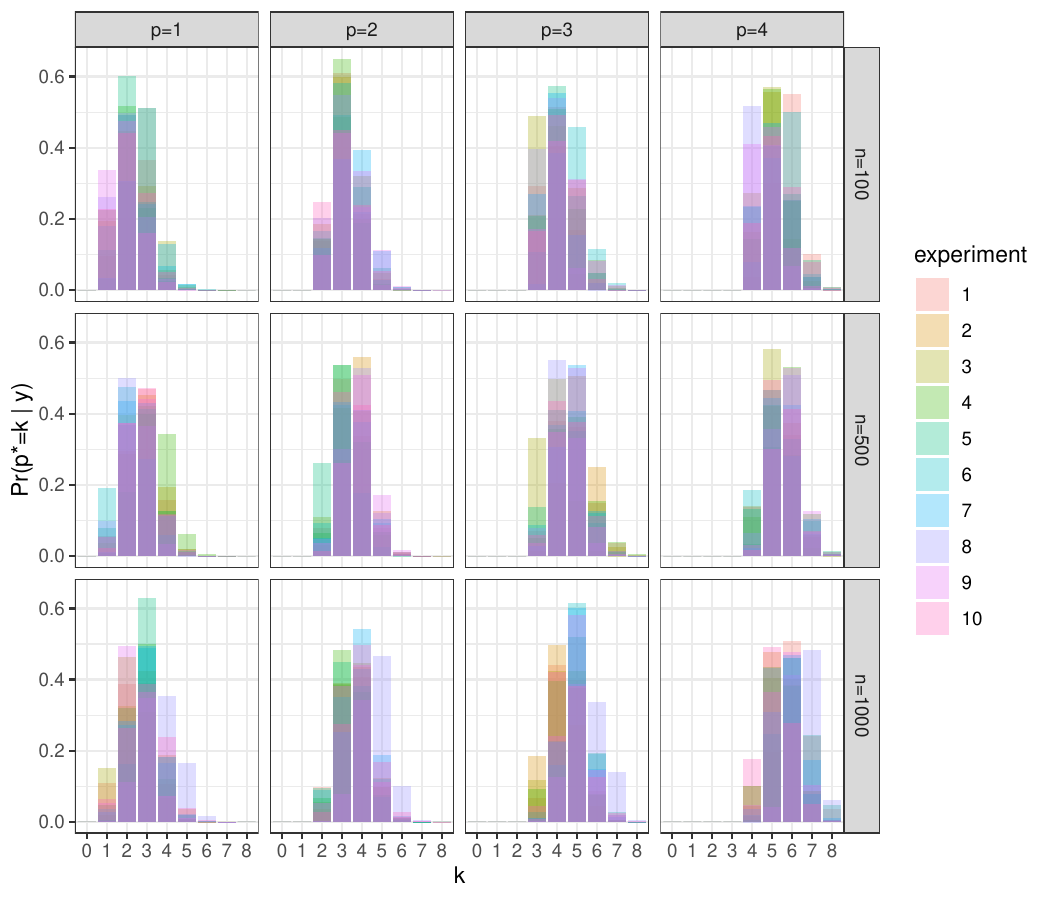}}
\caption{For $\beta=0.5$, overlaid posterior mass functions for the effective order $p^*$ from 10 experiments for each combination of $p \in \{1,2,3,4\}$ and (a) $m \in \{1,3,5,7\}$ with $n=1000$ or (b) $n \in \{100,500,1000\}$ with $m=3$.}
\label{fig:post_mass_05}
\end{center}
\end{figure}

\begin{figure}[!th]
\begin{center}
\subfloat[][]{\includegraphics[width=.725\textwidth]{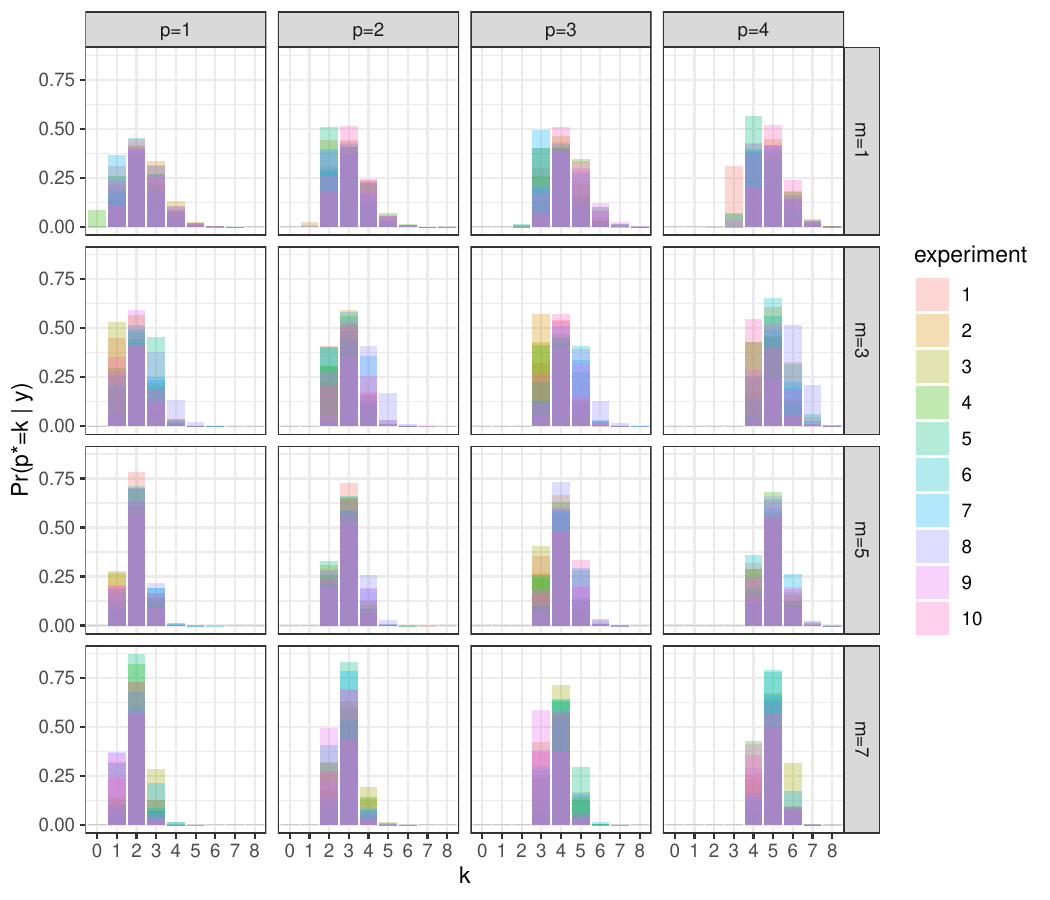}}\\
\subfloat[][]{\includegraphics[width=.725\textwidth]{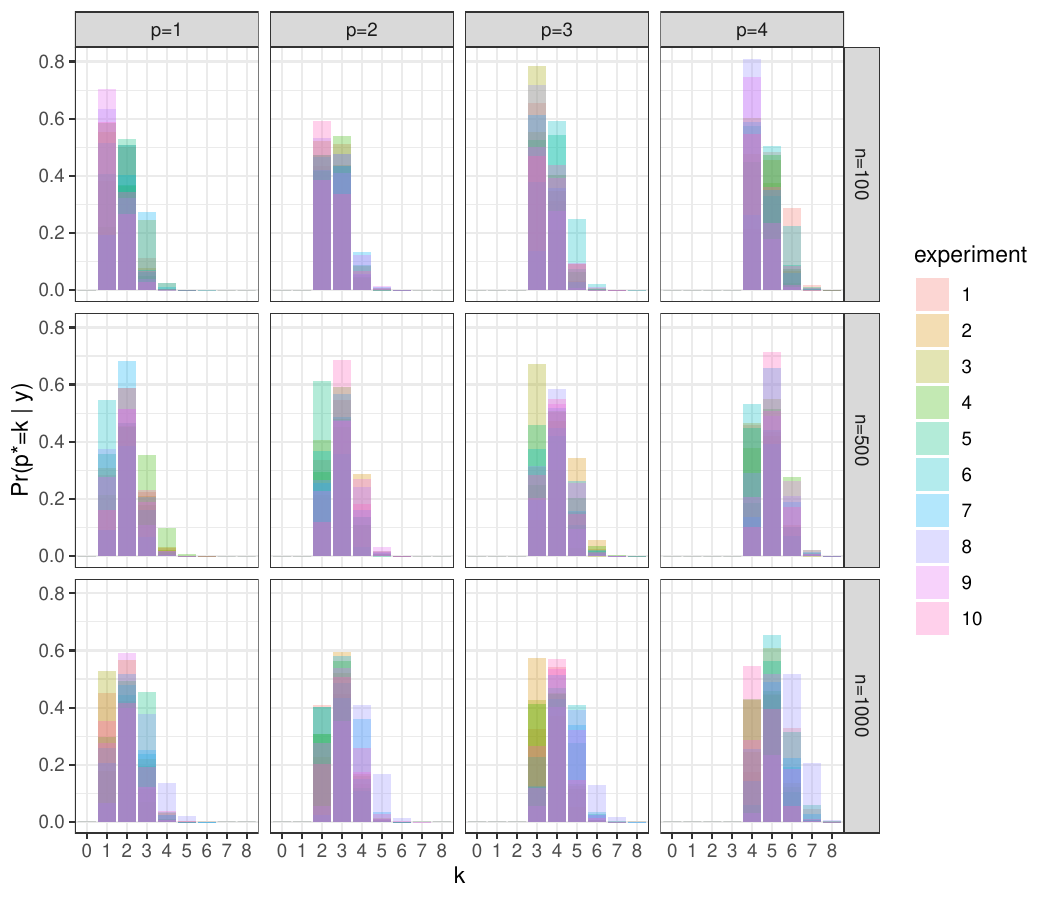}}
\caption{For $\beta=0.8$, overlaid posterior mass functions for the effective order $p^*$ from 10 experiments for each combination of $p \in \{1,2,3,4\}$ and (a) $m \in \{1,3,5,7\}$ with $n=1000$ or (b) $n \in \{100,500,1000\}$ with $m=3$.}
\label{fig:post_mass_08}
\end{center}
\end{figure}

\begin{figure}[!th]
\begin{center}
\subfloat[][]{\includegraphics[width=.725\textwidth]{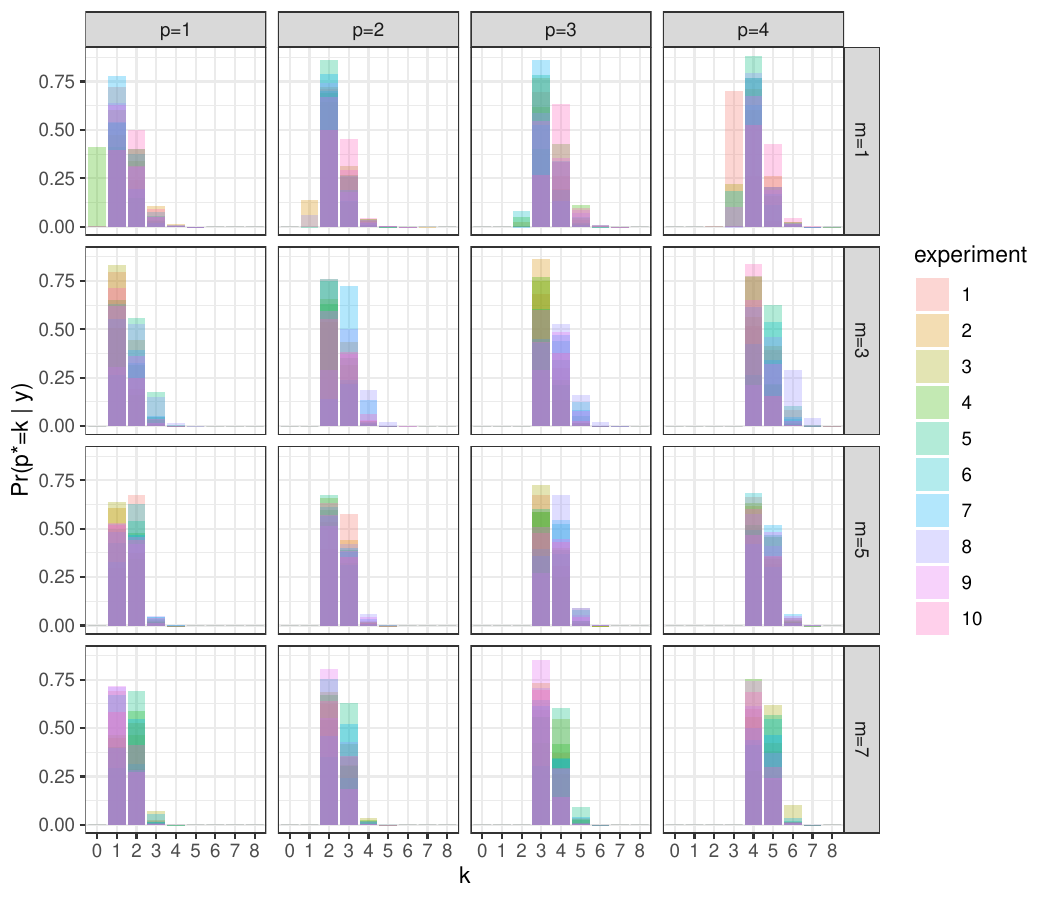}}\\
\subfloat[][]{\includegraphics[width=.725\textwidth]{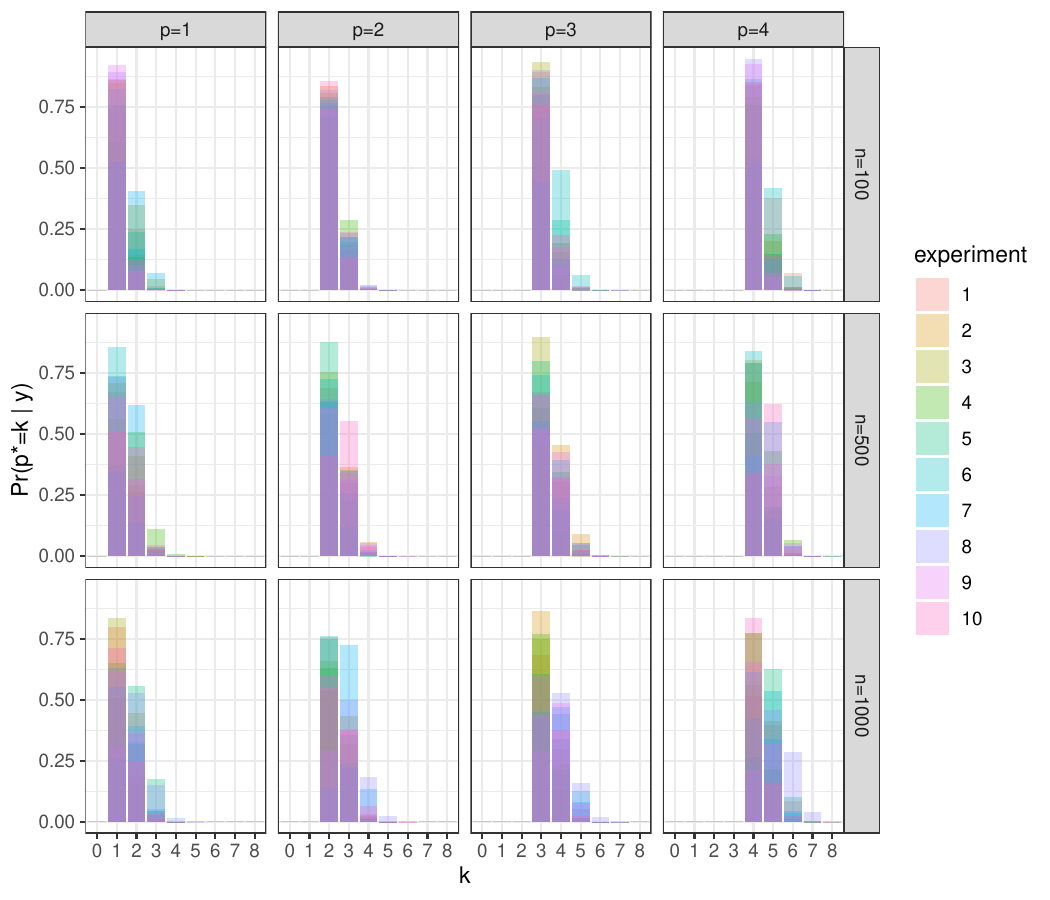}}
\caption{For $\beta=0.95$, overlaid posterior mass functions for the effective order $p^*$ from 10 experiments for each combination of $p \in \{1,2,3,4\}$ and (a) $m \in \{1,3,5,7\}$ with $n=1000$ or (b) $n \in \{100,500,1000\}$ with $m=3$.}
\label{fig:post_mass_095}
\end{center}
\end{figure}

\begin{figure}[!th]
\begin{center}
\subfloat[][]{\includegraphics[width=.725\textwidth]{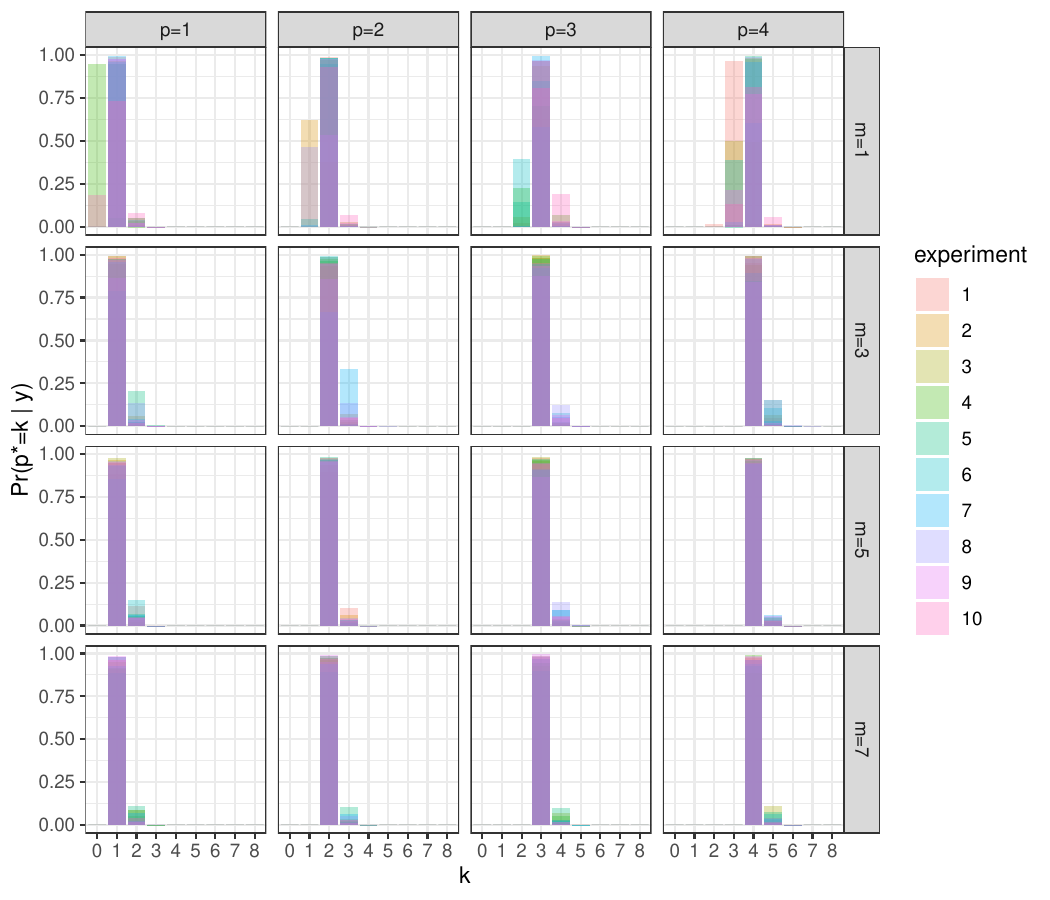}}\\
\subfloat[][]{\includegraphics[width=.725\textwidth]{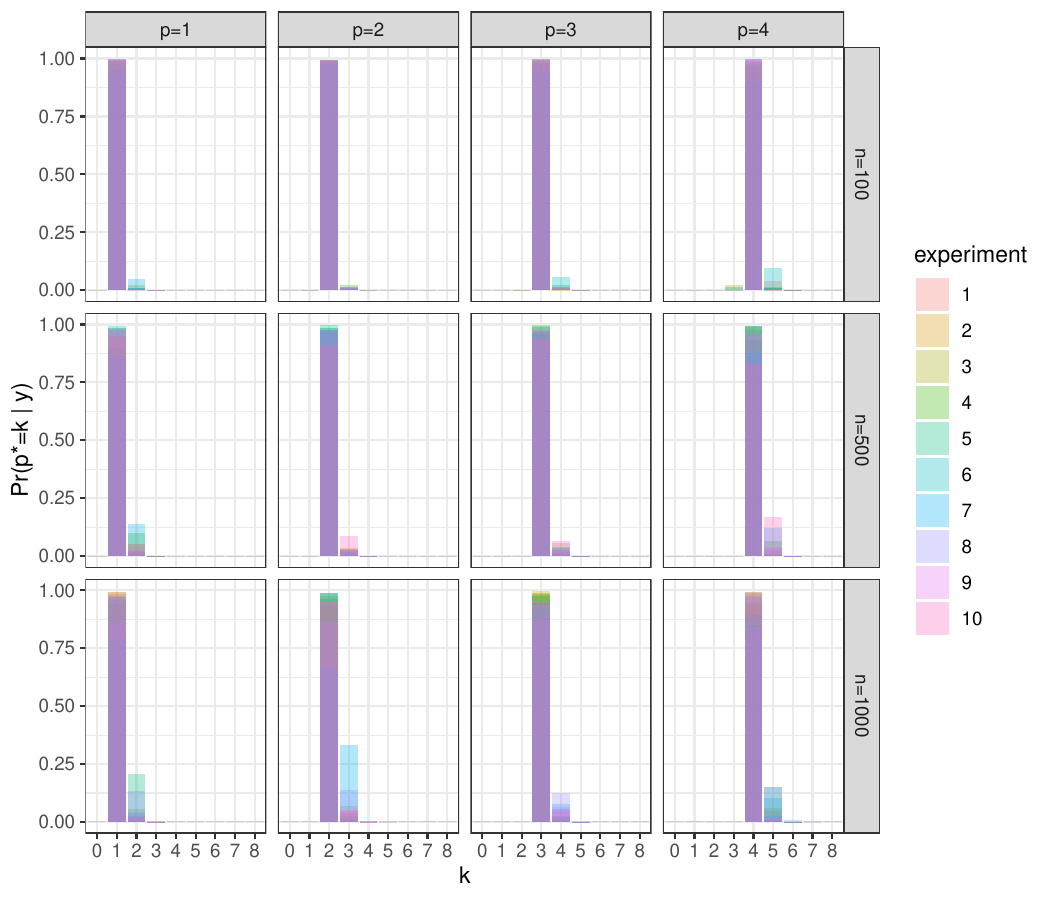}}
\caption{For $\beta=0.999$, overlaid posterior mass functions for the effective order $p^*$ from 10 experiments for each combination of $p \in \{1,2,3,4\}$ and (a) $m \in \{1,3,5,7\}$ with $n=1000$ or (b) $n \in \{100,500,1000\}$ with $m=3$.}
\label{fig:post_mass_0999}
\end{center}
\end{figure}

In the simulation experiments, we set $\beta=0.99$ in the truncation criterion
\begin{equation*}
\varepsilon=q_{m,n}(\beta)=\Phi^{-1}\{ (\beta^{1/m^2}+1) / 2 \} / \sqrt{n},
\end{equation*}
where the probability $\beta$ is equal to $\Pr\{\max_{i,j} | p_{s,ij} | < q_{m,n}(\beta) \}$ under the approximate posterior for $\matr{P}_s$ if the order of the process is less than $s$. In order to assess posterior sensitivity to our choice of truncation criterion, we considered four other choices for $\beta$, namely $\beta=0.5$, $\beta=0.8$, $\beta=0.95$ and $\beta=0.999$; the first two values were chosen to make the truncation criterion unjustifiably small whilst the second two were chosen to represent reasonable alternatives on either side of the chosen value of $\beta=0.99$. The results, presented in Figures~\ref{fig:post_mass_05}--\ref{fig:post_mass_0999}, reveal that the central tendency of the posterior for $p^*$ is largely insensitive to sensible variation in the value of $\beta$, with the posterior mode remaining equal to the known order of the process when $\beta=0.95$ and $\beta=0.999$. As expected, when $\beta=0.5$ or $\beta=0.8$, the truncation threshold is unreasonably small and so the posterior for the effective order $p^{\ast}$ tends to overestimate the known order, with the posterior mode typically being one or sometimes two (in the $\beta=0.5$ case) lags higher than $p$. For reasonable choices of $\beta$, the posterior spread is affected in the manner that would be expected, such that as $\beta$ increases from 0.95 through 0.99 to 0.999, the posterior for $p^{\ast}$ becomes more concentrated around the mode. Clearly, in the limit as $\beta \to 1$, the truncation criterion would become infinitely large and the posterior for $p^*$ would concentrate at zero.

\section{\label{sec:suppl_application}Application: understanding brain rhythms}

\subsection{Choice of hyperparameters in the prior}
We used the same choice of hyperparameters for our prior during the application to neural activity data as during the simulation experiments; see Section \ref{subsec:suppl_simulation_hyperparam}.

\subsection{Order determination}
\begin{figure}[!th]
\begin{center}
\subfloat[][]{\includegraphics[width=0.8\textwidth]{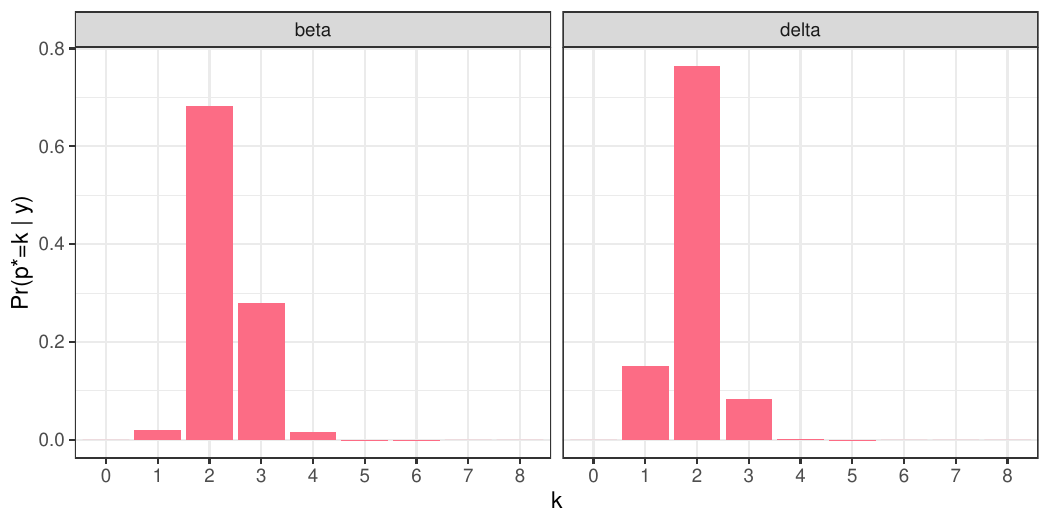}}\\
\subfloat[][]{\includegraphics[width=0.8\textwidth]{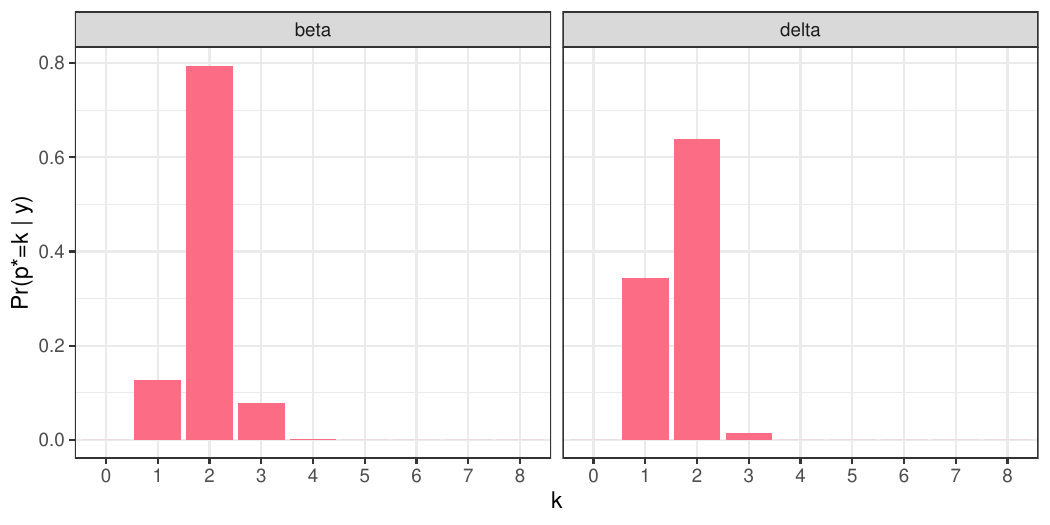}}\\
\subfloat[][]{\includegraphics[width=0.8\textwidth]{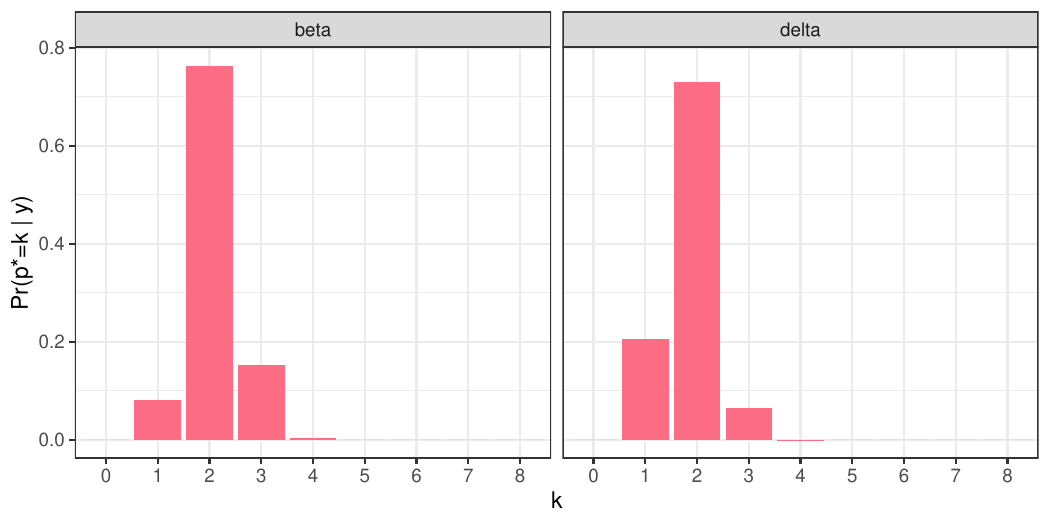}}
\caption{Posterior mass function for the order of the process for the beta (left) and delta (right) series for individuals (a) B, (b) C and (c) D.}
\label{fig:day_pmf}
\end{center}
\end{figure}

Figure 3 in the main manuscript presented the posterior mass function for the effective order $p^*$ of both the delta and beta series for individual A. For these data, the manuscript also provided a comparison to a frequentist analysis, with order selected by the AIC and SC. The corresponding plots for the other three individuals are presented in Figure~\ref{fig:day_pmf}. \label{pg:ed_3b}Similarly, the orders selected by the AIC and SC are displayed in Table~\ref{tab:order}. We note that the point estimates of the autoregressive coefficient matrices, on which the AIC and SC calculations are based, lay within the stationary region in all cases.

\begin{table}[bt]
\begin{center}
\begin{tabular}{ccccc}
\toprule
{\it Individual} &\multicolumn{2}{c}{\it Beta band} &\multicolumn{2}{c}{\it Delta band}\\
\cmidrule(lr){2-3} \cmidrule(lr){4-5}
                 &{\it AIC} &{\it SC}               &{\it AIC} &{\it SC}\\
\midrule
A                &2         &1                      &2         &1\\
B                &1         &1                      &1         &1\\
C                &2         &1                      &2         &1\\
D                &1         &1                      &1         &1\\
\bottomrule
\end{tabular}
\caption{Order selected by the AIC and SC for the data in the beta and delta bands for each individual.}\label{tab:order}
\end{center}
\end{table}

\subsection{Granger causality}
\begin{figure}[!t]
\begin{center}
\subfloat[][]{\includegraphics[width=1\textwidth]{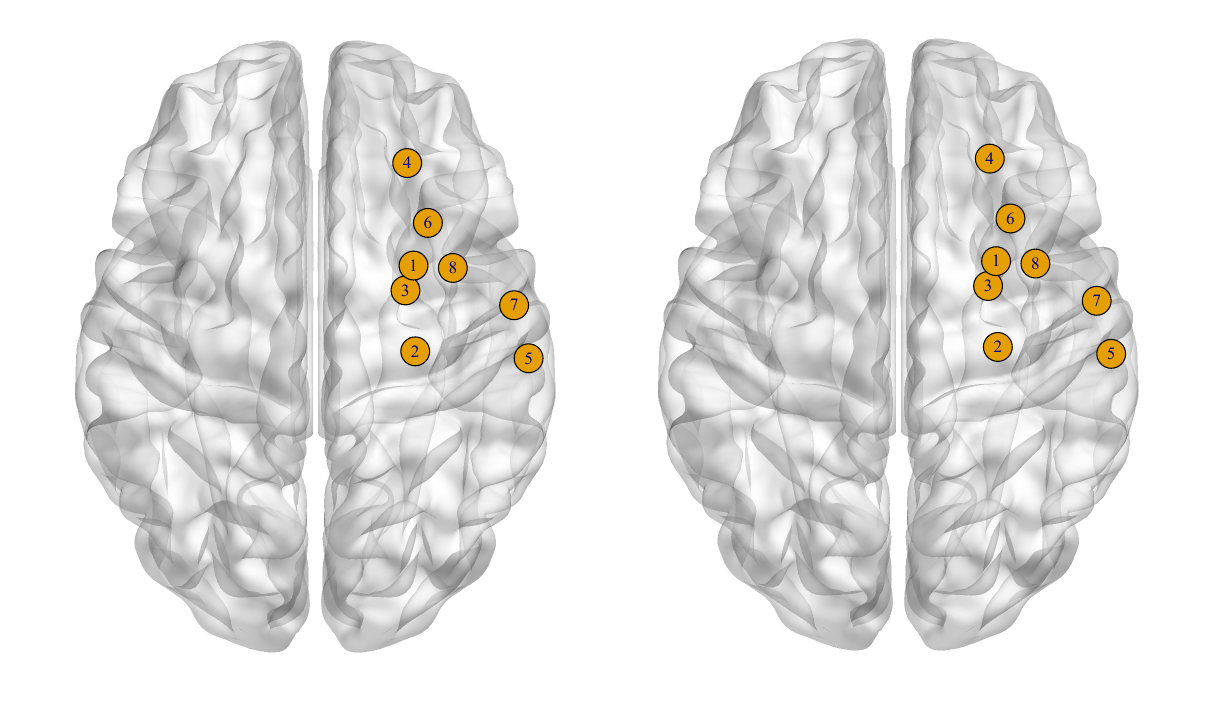}}\\
\subfloat[][]{\includegraphics[width=1\textwidth]{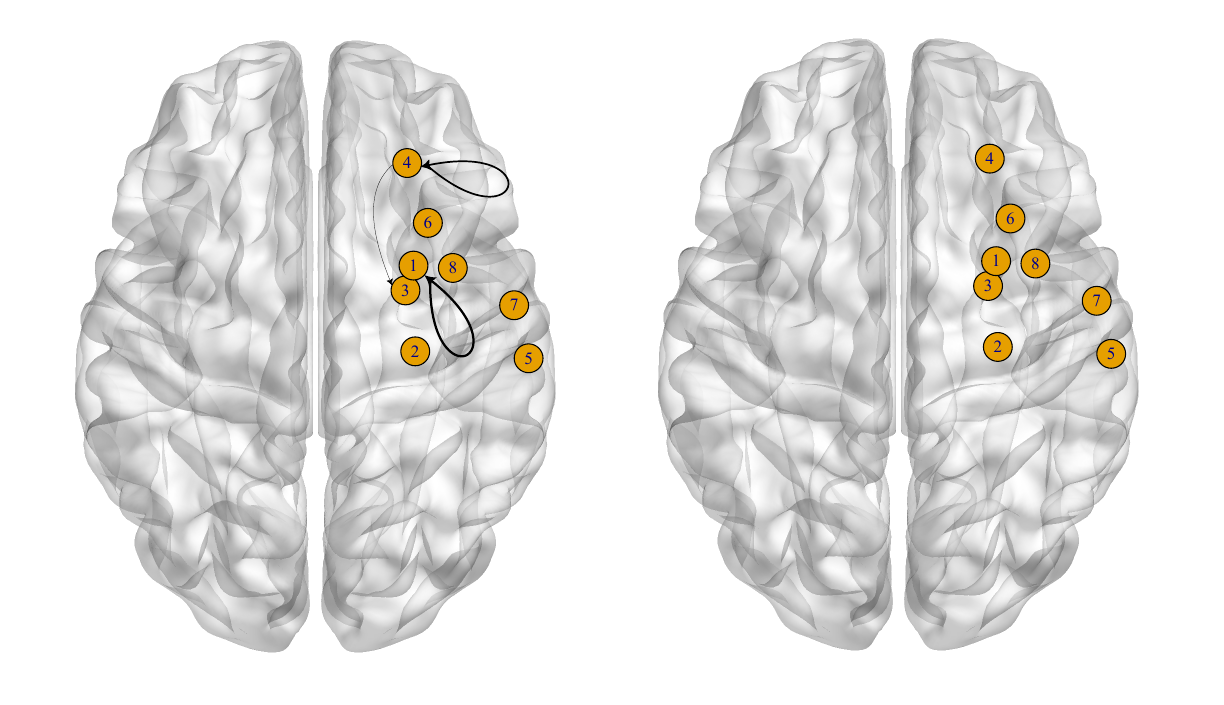}}
\caption{Granger causality plots of the posterior mean of the autoregressive coefficient matrices for the VAR process of individual B in the (a) beta and (b) delta band at lag 1 (left) and lag 2 (right), overlaid on glass brains showing the locations of the regions. In the beta series, there are no elements of the autoregressive coefficient matrices for which zero lies outside of the relevant 50\% equi-tailed Bayesian credible interval. Region names: 1 - right-putamen, 2 - right-hippocampus, 3 - right-amygdala, 4 - r.lateralorbitofrontal, 5 - r.middletemporal, 6 - r.temporalpole, 7 - r.superiortemporal, 8 - r.insula.}
\label{fig:beta_delta_1038_day_GC}
\end{center}
\end{figure}

\begin{figure}[!t]
\begin{center}
\subfloat[][]{\includegraphics[width=1\textwidth]{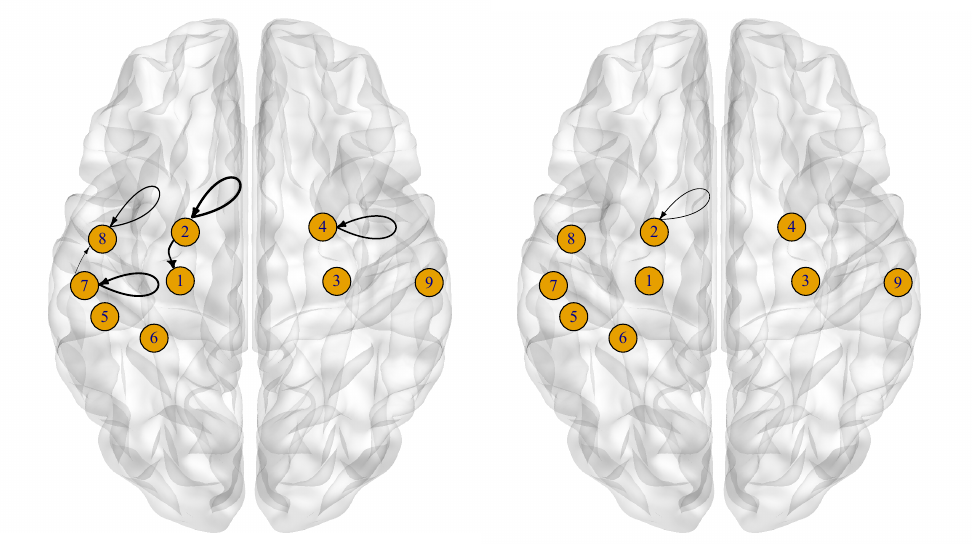}}\\
\subfloat[][]{\includegraphics[width=1\textwidth]{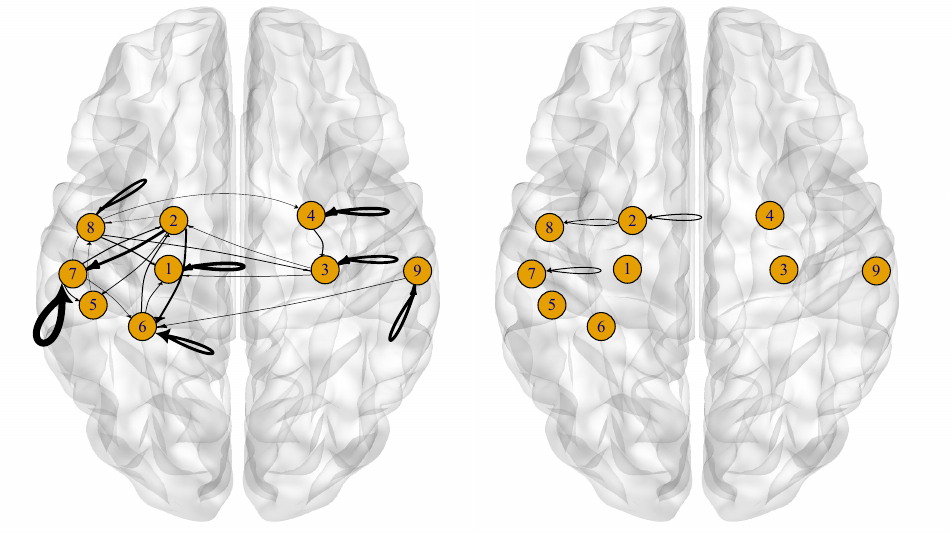}}
\caption{Granger causality plots of the posterior mean of the autoregressive coefficient matrices for the VAR process of individual C in the (a) beta and (b) delta band at lag 1 (left) and lag 2 (right), overlaid on glass brains showing the locations of the regions. Region names: 1 - left-hippocampus, 2 - left-amygdala, 3 - right-hippocampus, 4 - right-amygdala, 5 - l.inferiortemporal, 6 - l.fusiform, 7 - l.middletemporal, 8 - l.superiortemporal, 9 - r.middletemporal.}
\label{fig:beta_delta_909_day_GC}
\end{center}
\end{figure}

\begin{figure}[!t]
\begin{center}
\subfloat[][]{\includegraphics[width=1\textwidth]{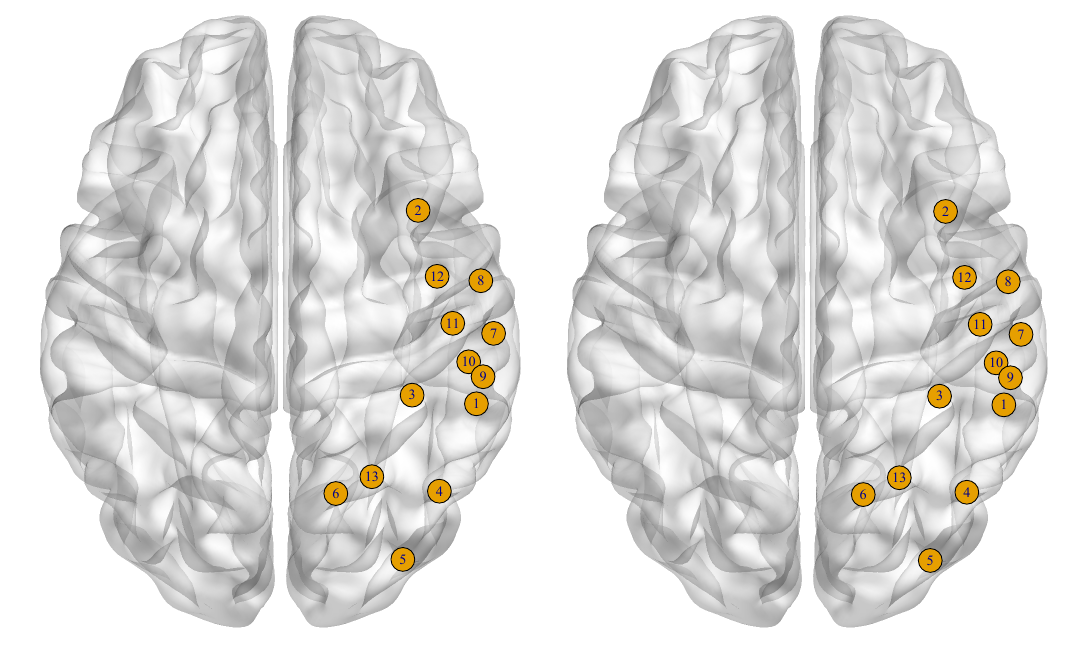}}\\
\subfloat[][]{\includegraphics[width=1\textwidth]{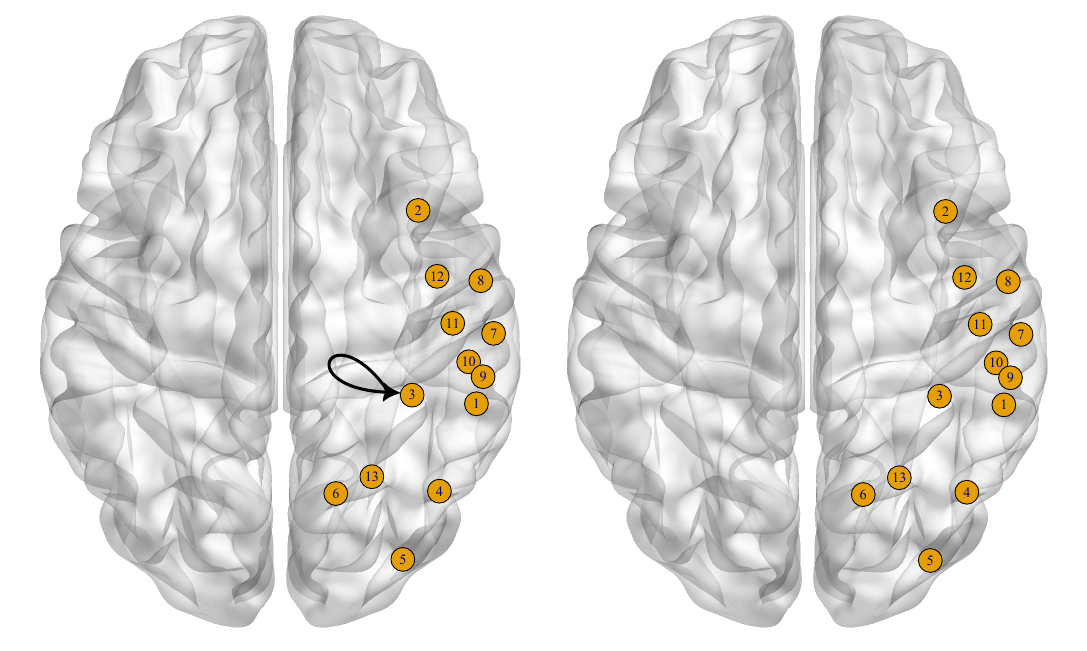}}
\caption{Granger causality plots of the posterior mean of the autoregressive coefficient matrices for the VAR process of individual D in the (a) beta and (b) delta band at lag 1 (left) and lag 2 (right). In the beta series, there are no elements of the autoregressive coefficient matrices for which zero lies outside of the relevant 50\% equi-tailed Bayesian credible interval. Region names: 1 - r.bankssts, 2 - r.caudalmiddlefrontal, 3 - r.fusiform, 4 - r.inferiorparietal, 5 - r.lateraloccipital, 6 - r.lingual, 7 - r.middletemporal, 8 - r.superiortemporal, 9 - r.supramarginal, 10 - r.inferiortemporal, 11 - r.postcentral, 12 - r.precentral, 13 - r.superiorparietal.}
\label{fig:beta_delta_934_day_GC}
\end{center}
\end{figure}

Figures 4 and 5 in the main manuscript display the Granger causality plots for individual A in the beta and delta bands. The corresponding plots for individuals B, C and D are presented in Figures~\ref{fig:beta_delta_1038_day_GC}--\ref{fig:beta_delta_934_day_GC}. In all cases the mode of the posterior for $p^\ast$ was equal to two and so we display graphs for lags 1 and 2 in each case.

\subsection{Latent moduli and periods}

\begin{figure}[!th]
\begin{center}
\subfloat[][]{\includegraphics[width=1\textwidth]{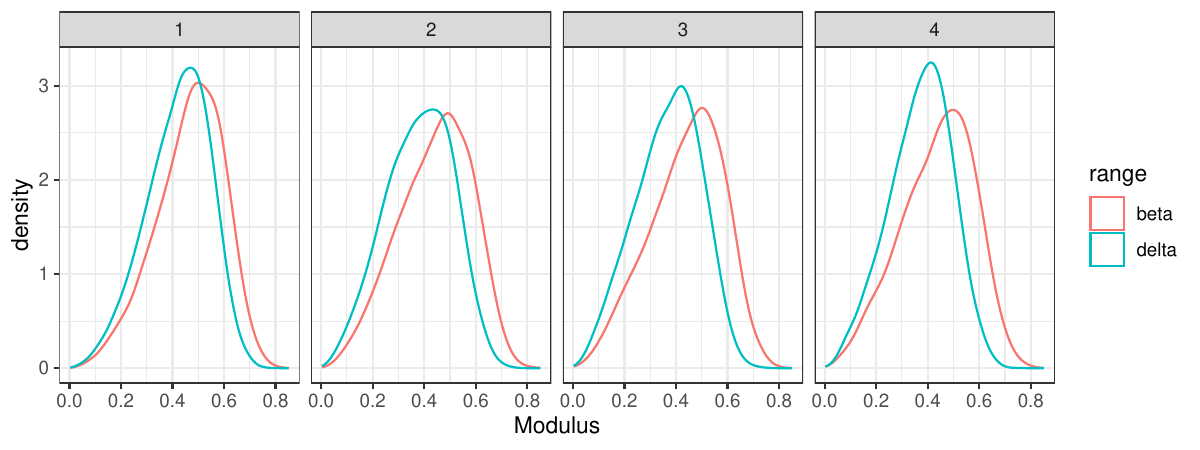}}\\
\subfloat[][]{\includegraphics[width=1\textwidth]{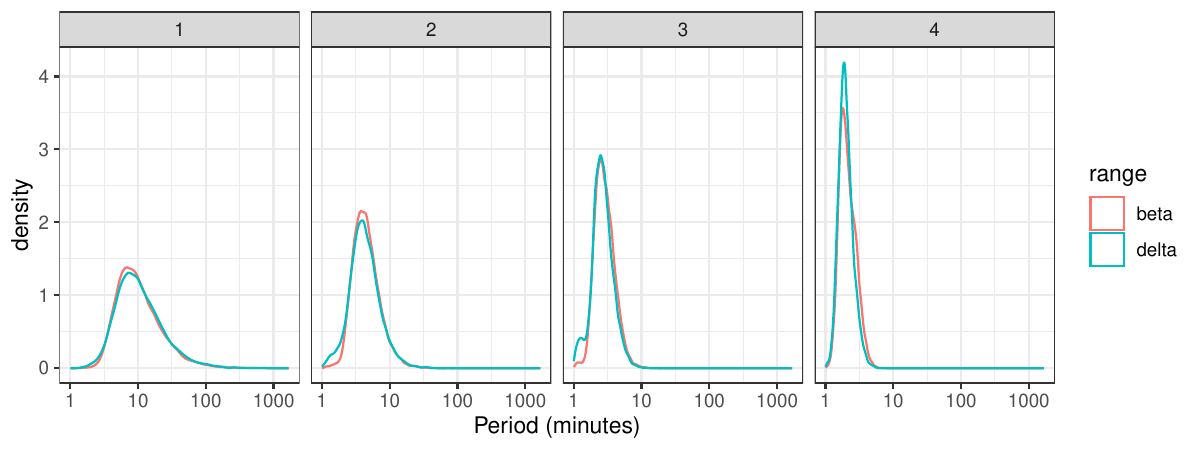}}
\caption{Posterior densities for the (a) moduli and (b) periods of the first four quasi-periodic series for individual  B.}
\label{fig:1038_moduli_periods_day}
\end{center}
\end{figure}

\begin{figure}[!th]
\begin{center}
\subfloat[][]{\includegraphics[width=1\textwidth]{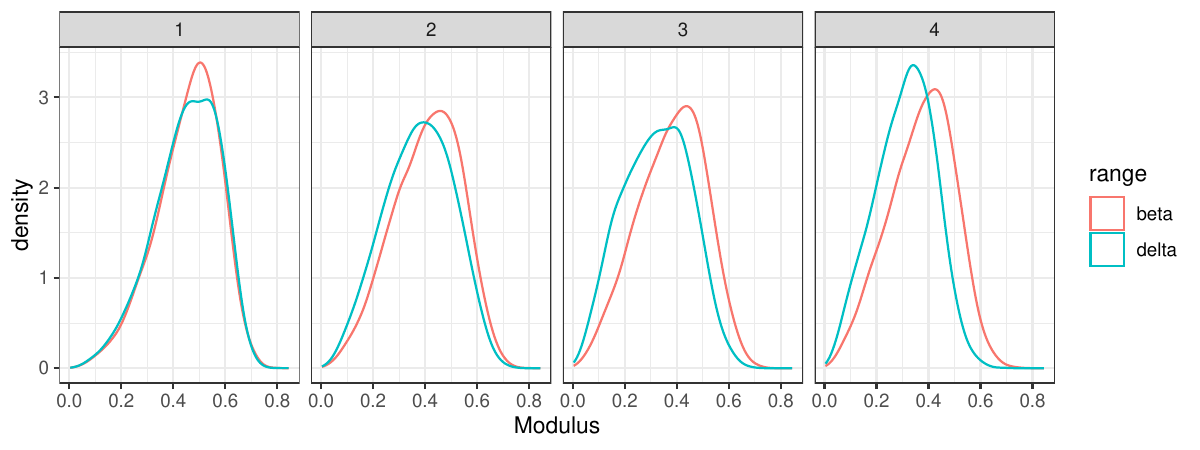}}\\
\subfloat[][]{\includegraphics[width=1\textwidth]{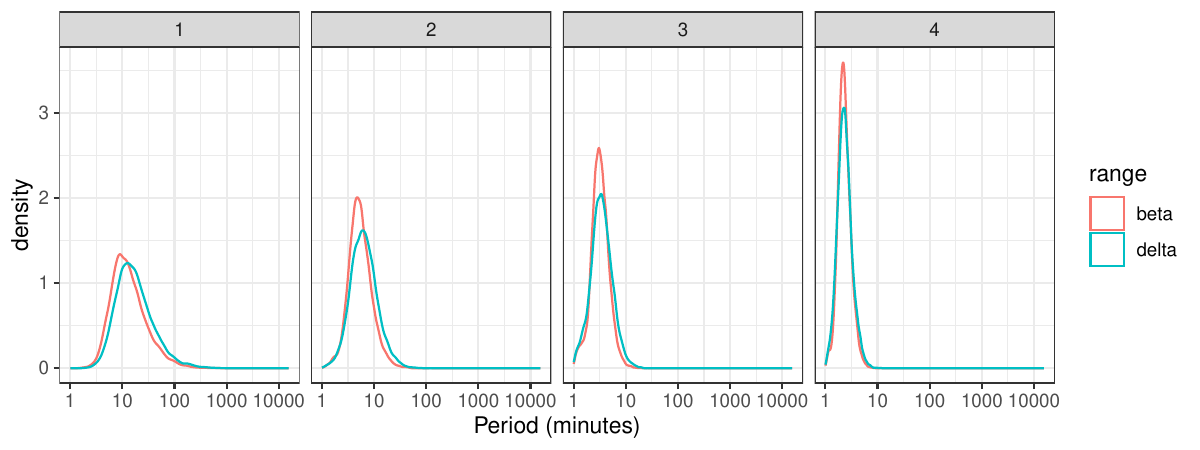}}
\caption{Posterior densities for the (a) moduli and (b) periods of the first four quasi-periodic series for individual C.}
\label{fig:909_moduli_periods_day}
\end{center}
\end{figure}

\begin{figure}[!th]
\begin{center}
\subfloat[][]{\includegraphics[width=1\textwidth]{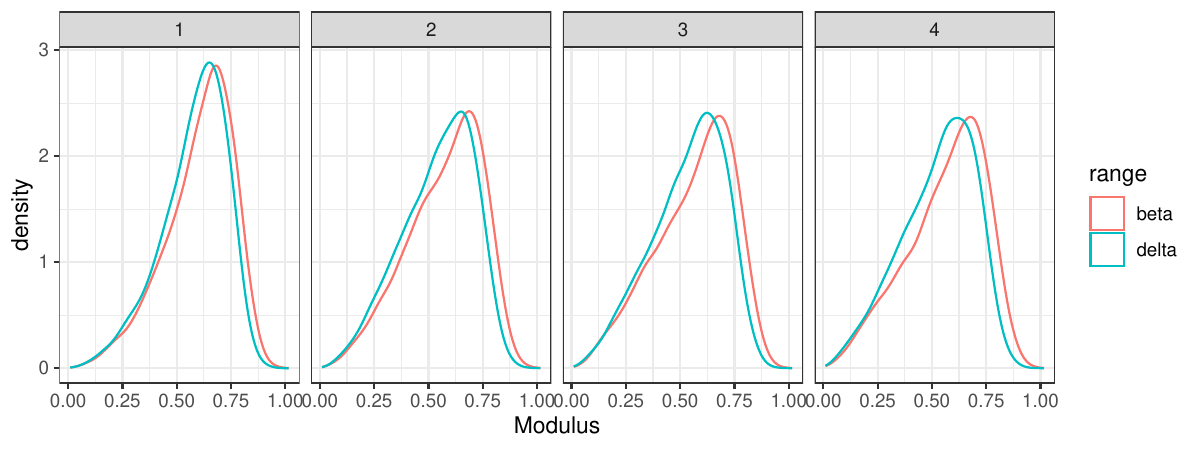}}\\
\subfloat[][]{\includegraphics[width=1\textwidth]{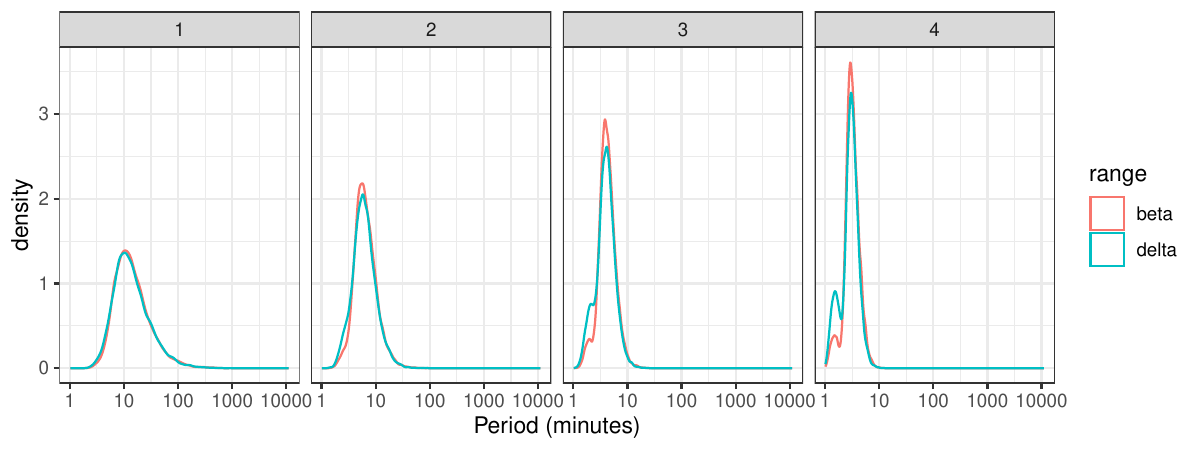}}
\caption{Posterior densities for the (a) moduli and (b) periods of the first four quasi-periodic series for individual D.}
\label{fig:934_moduli_periods_day}
\end{center}
\end{figure}

In the main manuscript, Figures 6 and 7 display the posterior densities for the moduli and periods of the first four quasi-periodic latent series for individual A. Corresponding plots for individuals B, C and D are presented in Figures~\ref{fig:1038_moduli_periods_day}--\ref{fig:934_moduli_periods_day}.

\bibliographystyle{chicago}
\bibliography{references.bib}